\documentclass[useAMS,usenatbib,usedcolumn,usegraphicx]{mn2e}
\usepackage{xspace}
\usepackage{color,epsfig,rotating,amssymb,longtable}
\usepackage{array,colortbl}
\usepackage[colorlinks=true,linkcolor=magenta,citecolor=blue,breaklinks=true]{hyperref}

\usepackage{ifpdf}

\ifpdf
\usepackage{epstopdf}
\else
\usepackage[dvips]{graphicx}
\usepackage{lscape}
\fi

\newcommand{\HII}{H{\sc ii}\xspace}
\newcommand{\Ha}{\ensuremath{{\rm H\alpha}}\xspace}
\newcommand{\Msun}{\ensuremath{{\rm M_\odot}}\xspace}

\newcommand{\OIII}{O{\sc iii}}
\newcommand{\OII}{O{\sc ii}}
\newcommand{\OI}{O{\sc i}}
\newcommand{\SIII}{S{\sc iii}}
\newcommand{\SII}{S{\sc ii}}
\newcommand{\NII}{N{\sc ii}}
\newcommand{\NI}{N{\sc i}}
\newcommand{\HeI}{He{\sc i}}
\newcommand{\HeII}{He{\sc ii}}
\newcommand{\NeIII}{Ne{\sc iii}}
\newcommand{\FeIII}{Fe{\sc iii}}

\newcommand{\ArIII}{Ar{\sc iii}}

\def\seis{SDSS J165712.75+321141.4\xspace}  

\def\seisc{SDSS J1657\xspace}  

\usepackage{color}

\definecolor{dgreen}{rgb}{0,.5,.1} 
\definecolor{pink}{rgb}{.9,.2,.5}  
\definecolor{orange}{rgb}{.9,.4,0} 
\definecolor{darkred}{rgb}{.545,0.0,.0}
\title[Abundance of multiple knots in \seisc]
{Abundance determination of multiple star-forming regions in the \HII galaxy \seis}

\author[G. F. H\"agele et al.]
       {Guillermo F. H\"agele$^{1,2}$\thanks{CONICET, Argentina; e-mail:
       guille.hagele@uam.es, ghagele@fcaglp.edu.ar},
       Rub\'en Garc\'{\i}a-Benito$^{3,1}$, Enrique P{\'e}rez-Montero$^{4}$,
\newauthor \'Angeles I. D\'{\i}az$^{1}$, M\'onica V. Cardaci$^{1,2}$,
       Ver\'onica Firpo$^{2}$, Elena Terlevich$^{5}$\thanks{Visiting
       astronomer at IoA, University of Cambridge, UK} 
\newauthor and Roberto Terlevich$^{5}$\thanks{Research Affiliate, IoA, University of Cambridge, UK} \\ 
$^{1}$ Departamento de F\'{\i}sica Te\'orica, M\'odulo 15, Universidad Aut\'onoma de
Madrid, 28049 Madrid, Spain\\ 
$^{2}$ Facultad de Cs.\ Astron\'omicas y Geof\'isicas, Universidad Nacional de La
Plata, Paseo del Bosque s/n, 1900 La Plata, Argentina \\ 
$^{3}$ Kavli Institute of Astronomy and Astrophysics, Peking University,
       100871, Beijing, China\\ 
$^{4}$ Instituto de Astrof\'\i sica de Andaluc\'\i a, CSIC, Apdo. 3004, 18080, Granada, Spain.\\
$^{5}$ Instituto Nacional de Astrof\'isica, \'Optica y Electr\'onica,
Tonantzintla, Apdo. Postal 51, 72000 Puebla, M\'exico\\ 
}

\begin{document}

\maketitle

\begin{abstract}

We analyze high signal-to-noise spectrophotometric observations acquired
simultaneously with TWIN, a double-arm spectrograph, from 3400 to 10400\,\AA\ 
of three star-forming regions in the \HII galaxy
\seis. 
We have measured four line temperatures: T$_e$([\OIII]), T$_e$([\SIII]),
T$_e$([\OII]), and T$_e$([\SII]), with high precision, rms errors of 
order  2\%, 5\%, 6\% and 6\%, respectively, for the brightest region, and
slightly worse for the other two. The 
temperature measurements allowed the direct derivation of ionic abundances 
of oxygen, sulphur, nitrogen, neon and argon. 

{We have computed CLOUDY tailor-made models which reproduce the O$^{2+}$
measured thermal and ionic structures within the errors in the three knots,
with deviations of only 0.1 dex in the case of O$^+$ and S$^{2+}$ ionic
abundances. In the case of the electron temperature and the ionic abundances
of S$^+$/H$^+$, we find major discrepancies which could be consequence of the
presence of colder diffuse gas.} 
The star formation history  derived using STARLIGHT shows
a similar age distribution of the ionizing population among the three star-forming regions.
This fact suggests a similar evolutionary history which is probably related to
a process of interaction with a companion galaxy that triggered the star
formation in the different regions almost at the same time.  
The hardness of the radiation field mapped through the use of the
softness parameter $\eta$ is the same within the observational errors for all three regions,
implying that the equivalent effective temperature of the
radiation fields are very similar
for all the studied regions of the galaxy, in spite of some small
differences in the ionization state of different elements.

Regarding the kinematics of the galaxy, the gas rotation curve shows a deviation from
the circular motion probably due either to an interaction process or 
related to an expanding bubble or shell of ionized gas approaching us. 
A dynamical mass of 2.5\,$\times$\,10$^{10}$\,\Msun\ is derived 
from the rotation curve.

\end{abstract}


\section{Introduction}
\label{secIntro}

Star formation is an ongoing process in the local universe, with observed
rates of the order of 10$^{-2}$ M$_{\odot}$ yr$^{-1}$ Mpc$^{-3}$
\citep{Madau+96}. Most of the light and metals are produced in the most massive
among the newly formed stars. The most extreme regions forming massive stars
are often referred to as starbursts. In the local universe they account for
about a quarter of all star formation \citep{Heckman97}, and this fraction may have been
larger in the younger universe. The origin of the term ``starburst''
\citep[coined as ``starburst nuclei" by][]{Weedman+81} dates
back to the early observations of dust-obscured star-forming regions in the
centres of nearby galaxies at the end of the seventies and beginning of the
eighties, but the basic concept extends further back
\citep[\textit{e.g.},][]{Hodge69b,Searle+73}.

The level of intensity of a starburst is highly variable. According to
\cite{Terlevich97}, in a starburst galaxy the energy output of the starburst
($L_{SB}$) is much larger than the one coming from the rest of the galaxy
($L_{G}$), a galaxy with $L_{SB}$ $\sim$ $L_{G}$ is a galaxy with starbursts,
and in a normal galaxy $L_{SB}$ $\ll$ $L_{G}$. This classification shows the
variety of environments of the bursts. It is clear that the visibility of the
burst depends not only on its intensity but also on its environment. 
\cite{Terlevich97} also proposed a division in phases of the starburst. The
first one, the nebular phase, is characterized by the presence of strong
emission lines from gas photoionized by young massive stars, with an age of
less than 10 Myr. The early continuum phase goes from 10 to 100 Myr, when some 
Balmer lines appear in absorption and others in emission. Finally, the late
continuum phase, is when  only some weak emission lines appear in the
spectrum. The \HII galaxies are typical examples of the first phase.

\HII galaxies are gas-rich dwarf galaxies experiencing a violent star
formation period which dominates the optical spectrum of the host galaxy. They
have one of the highest intensity levels of star forming activity. In general,
these galaxies have a central region which contains one or more star forming
knots, with a diameter of several hundred parsecs with high surface
brightness, and a low luminosity underlying galaxy (M$_{V}$ $\geq$ -17). The
activity of the star formation episodes can not be sustained continuously for
long periods of time, since the central region can not have enough gas to fuel
these processes for longer than 10$^{9}$ years and to match the gas content and
metallicity with theoretical considerations \citep{Thuan+04}. 

Spectroscopically, \HII galaxies are essentially identical to the giant \HII
regions found in nearby irregular and late-type galaxies. The correlation
among structural parameters (H$\beta$ luminosity, velocity dispersion, line
widths) and between these parameters and chemical composition
\citep{Terlevich+81} favours the interpretation of \HII galaxies as giant \HII
regions in distant dwarf irregular galaxies similar to the ones found nearby
\citep{Melnick+85}. 

Other important characteristic of \HII galaxies is their low metallicity
\citep[Z$_{\odot}$/50 $\leq$ Z $\leq$ Z$_{\odot}$/3; ][]{Kunth+83}. The fact
that \HII galaxies are metal-poor  and very blue objects seems to suggest that 
they are young. Nevertheless, there is evidence which indicates the
presence of populations older than the ones in the starburst. This is seen in
the behaviour of the surface brightness profile which is exponential in the
external zones, or in the colour index, which turns redder in V-R and
V-I \citep{Telles+97}.  IZw18 in particular, was considered as the best
candidate for a truly young galaxy. Early studies of the stellar population of
IZw18 did 
not reveal any old population \citep{Hunter+95}. This contradicted some models
which predict that during a starburst, the heavy elements produced by the
massive stars are ejected with high velocities into a hot phase, leaving the
starburst region without immediate contribution to the enrichment of the
interstellar medium \citep{Tenorio-Tagle96}. In this scenario, the metals
observed now would have their origin in a previous star formation event, and
an underlying old stellar population would be expected. 
In fact, \cite{Garnett+97} attributed the high carbon abundance that they found 
in HST spectroscopy of IZw~18 as evidence for the presence of an old stellar population.
In agreement with this result, using
HST archive data \citep{Aloisi+99} showed that stars older than 1 Gyr must
be present in IZw~18. Moreover, studies of the resolved stellar population in the near
infrared with NICMOS \citep{Ostlin00} found also that while the near infrared
colour-magnitude diagram was dominated by stars 10-20 Myr old, the presence of
numerous AGB stars require an age of at least 10$^{8}$
years. \cite{Legrand+00} modelled the relative abundance of metals in IZw~18
and concluded that,  in addition to the present burst of star formation,  a
low star formation rate extended over a long period of time was necessary to
account for the observed values. 

In recent years, with the development of the Integral Field Unit (IFU)
instruments to perform 3D spectroscopy,  works that require a spatial
coverage to study extended galactic or extra galactic star-forming regions
 have been mainly focused on the use of this technique
\citep[see for
  example]{Relano+10,Cairos+10,Monreal-Ibero+10,Rosales-Ortega+10,Perez-Gallego+10,Garcia-Benito+10,Sanchez+10,Perez-Montero+11}. However,
medium or high dispersion  slit spectroscopy are a better option for
spectrophotometry, or when the object is very compact, or even extended but
with few star-forming knots. This is also the case when good spatial and
spectral resolution and  simultaneous wide spectral coverage are required
\citep[see for
  example]{Diaz+07,Cumming+08,Firpo+10,Hagele+06,Hagele+07,Hagele+08,HagelePhD,Hagele+09,Hagele+10,Perez-Montero+09,Lopez-Sanchez+09,Lopez-Sanchez+10a,Lopez-Sanchez+10b,Lopez-Sanchez+10c,Firpo+11a}. 

In this paper we present simultaneous blue and red long-slit observations obtained with
the double arm TWIN spectrograph at the 3.5m telescope of Calar Alto of the three brightest
star-forming knots of the \HII galaxy \seis. This is  part of a project to
obtain a top quality spectrophotometric data base to  determine ionized gas
parameters which are indispensable to critically test photoionization models
and to explore  discrepancies between models and observations. In 
Section 2 we show the details of the observations and data
reduction. Section 3 presents the derived physical characteristics of the
regions, including the electron temperature for four different
species. Section 4 is devoted to the discussion of these 
results, and finally the summary and conclusions  are presented in
Section 5.

%
%

\begin{table*}
\centering
\caption[]{Right ascension, declination, redshift and SDSS photometric
  magnitudes of the observed knots obtained using the SDSS explore tools$^a$.}
\label{obj}
\begin{tabular}{@{}cccccccrrcc@{}}
\hline
 Object  ID     & hereafter ID & Knot & RA & Dec &  redshift  & u & g & r & i   &  z \\
(spSpec SDSS)   &              &      &    &     &            &   &   &   &     &    \\

\hline
\seis  &  \seisc             &  A & 16$^h$\,57$^m$\,12\fs75 & 32$^{\circ}$\,11\arcmin\,41\farcs42 &  0.038 & 17.63 & 17.01 & 17.25 & 17.14 & 17.16 \\
(spSpec-52791-1176-591) &    &  B & 16$^h$\,57$^m$\,12\fs26 & 32$^{\circ}$\,11\arcmin\,43\farcs20 &   ---  & 20.34 & 19.76 & 20.65 & 20.72 & 20.27  \\
                        &    &  C & 16$^h$\,57$^m$\,13\fs58 & 32$^{\circ}$\,11\arcmin\,40\farcs09 &   ---  & 19.28 & 18.75 & 19.52 & 19.40 & 19.17  \\
\hline
\multicolumn{10}{l}{$^a$http://cas.sdss.org/astro/en/tools/explore/obj.asp}
\end{tabular}
\end{table*}


\section{Observations and data reduction}
\label{secObs}

\subsection{Object selection}

Using the implementation of the SDSS database in the INAOE Virtual Observatory
superserver\footnote{http://astro.inaoep.mx/en/observatories/virtual/}, we
selected the brightest nearby 
narrow emission line galaxies with very strong lines and large equivalent
widths of  H$\alpha$  from the whole SDSS data release available at the time of
planning the observations. These preliminary lists were then processed using BPT
\citep*{Baldwin+81} diagnostic diagrams to remove AGN-like objects. The final
list  consisted of about 10500 bonafide bright 
\HII galaxies. They show spectral properties indicating a wide range of
gaseous abundances and ages of the underlying stellar populations
\citep{jesustesis}. From this list, the final set was
selected by further restricting the sample to the largest H$\alpha$ flux and
highest signal-to-noise ratio objects \citep[for a complete description of the 
selection criteria see][hereafter Paper I]{Hagele+06}. Of the selected sample, we chose 
\seis to be observed at the allocated time. For simplicity, we will call the galaxy \seisc in what remains of the paper.

Some general characteristics of the knots of \seisc collected from the SDSS
web page are listed in Table \ref{obj}.


\begin{figure}
\centering
  \includegraphics[width=0.45\textwidth]{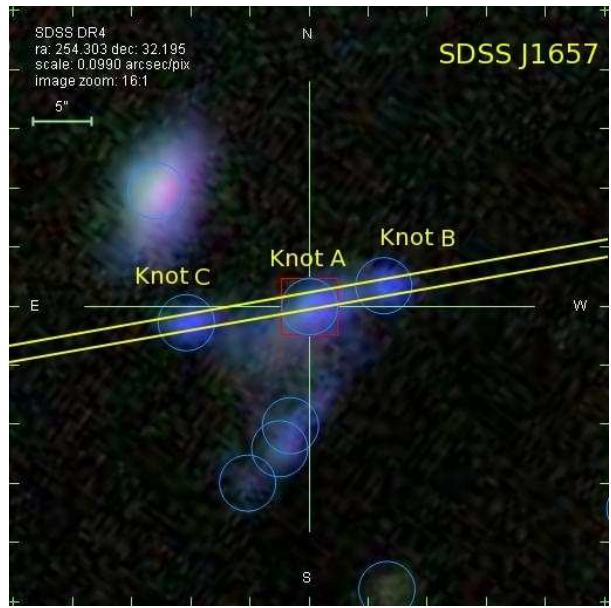}
  \caption[]{False colour image of \seisc\ with the slit position and the
    adopted knot names superimposed. This image was obtained using the SDSS
  explore tools. Circles and squares represent the photometric and
  spectroscopic SDSS targets, respectively. The scale is
  782\,pc\,arcsec$^{-1}$, at the adopted distance for \seisc. 
  [{\it See the electronic edition of the Journal for a   
  colour version of this figure.}]}
\label{imgalknots}
\end{figure}


\subsection{Observations}

Blue and red spectra were obtained simultaneously using the double beam
Cassegrain Twin Spectrograph (TWIN) mounted on the 3.5m telescope of the Calar
Alto Observatory at the Centro Astron\'omico Hispano Alem\'an (CAHA),
Spain. These observations were part of a four night observing run in 2006 June
and they were acquired under excellent seeing and photometric conditions
\cite[for details see;][hereinafter Paper II]{Hagele+08}. 
The blue arm covers the wavelength range 3400-5700\,\AA\ (centred at
$\lambda_c$\,=\,4550\,\AA), giving a spectral dispersion of
1.09\,\AA\,pixel$^{-1}$ (R\,$\simeq$\,4170). On the red arm, the spectral
range covers from 5800 to 10400\,\AA\ ($\lambda_c$\,=\,8100\,\AA) with a
spectral dispersion of 2.42\,\AA\,pixel$^{-1}$ (R\,$\simeq$\,3350). The
slit width was $\sim$\,1.2\,arcsec. The pixel
size for this set-up configuration is 0.56\,arcsec for both spectral
ranges. 
The target was observed at paralactic angle to avoid effects of
differential refraction in the UV. As it can be seen in Fig.\
\ref{imgalknots}, the three main knots of \seisc\ are 
almost perfectly aligned along the paralactic angle. The instrumental
configuration, summarized in Table \ref{config}, covers the whole spectrum
from 3400 to 10400\,\AA\ (with a gap between 5700 and 5800\,\AA) providing 
 a moderate spectral resolution.  This spectral coverage
guarantees the simultaneous detection of the nebular lines from [O{\sc
ii}]\,$\lambda\lambda$\,3727,29 to [S{\sc
iii}]\,$\lambda\lambda$\,9069,9532\,\AA\ at both ends of the spectrum, in the
very same region of the galaxy, with a good signal-to-noise ratio (S/N) that 
allows the measurement of the weak auroral lines. Typical values of S/N
 are  $\sim$\,60 for [O{\sc iii}]\,$\lambda$\,4363 and
$\sim$\,20 for [S{\sc ii}]\,$\lambda$\,4068 (see Table 4 of Paper
II). 

\begin{table}
\centering
\caption{CAHA instrumental configuration.}
\label{config}
\begin{tabular}{@{}lcccc@{}}
\hline
& Spectral range  &       Disp.        & R$^a_{\textrm{FWHM}}$     & Spatial res.          \\
& \multicolumn{1}{c}{(\AA)} &  (\AA\,px$^{-1}$)   &    & ($^{\prime\prime}$\,px$^{-1}$)    \\
\hline
Blue & 3400-5700       &       1.09          &  1420    &   0.56                    \\
Red  & 5800-10400      &       2.42          &  1160    &   0.56                     \\
\hline
\multicolumn{4}{l}{$^a$R$_{\textrm{FWHM}}$\,=\,$\lambda$/$\Delta\lambda_{\textrm{FWHM}}$}
\end{tabular}
\end{table}

\subsection{Data reduction}

Several bias and sky flat field frames were taken at the beginning and at the
end of the night in both arms. In addition, two lamp flat fields and one He-Ar 
calibration lamp exposures were performed at each telescope position. The
images were processed and analysed with IRAF\footnote{IRAF: the Image
  Reduction and Analysis Facility is distributed by the National Optical
  Astronomy Observatories, which is operated by the Association of
  Universities for Research in Astronomy, Inc. (AURA) under cooperative
  agreement with the National Science Foundation (NSF).} routines in the usual
manner. This procedure includes the removal of cosmic rays, bias subtraction,
division by a normalised flat field, and wavelength calibration. To finish,
the spectra are corrected for atmospheric extinction and
flux-calibrated. Four standard star observations were performed each night at
the same time for both arms, allowing a good spectrophotometric calibration
with an estimated rms error of about 3\%. Further details concerning each of
these steps can be found in Paper II.

\begin{figure}
\centering
  \includegraphics[width=0.46\textwidth]{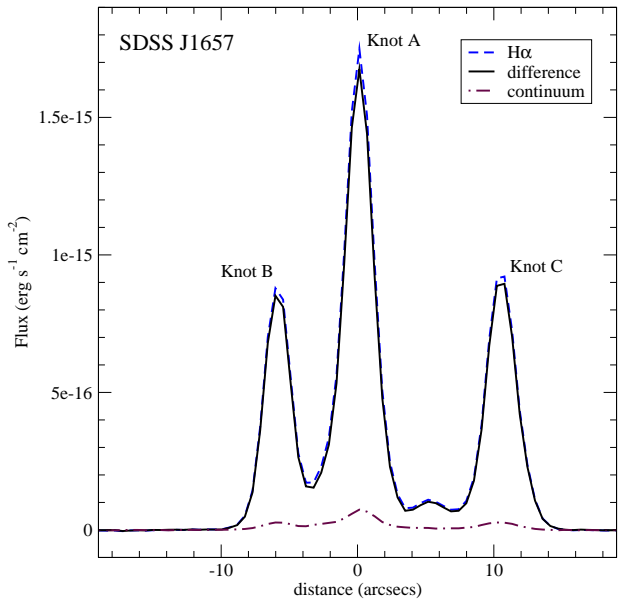}
  \caption[]{Spatial profile of the light distribution along the slit for the
    observed H$\alpha$ emission. The profiles correspond to
    line+continuum (dashed line), continuum (dashed-dotted line) and the
    difference between them (solid line), representing the pure emission from
    H$\alpha$.} 
\label{profiles}
\end{figure}

Fig.\ \ref{profiles} shows the spatial distribution of the H$\alpha$
flux and the continuum along the slit for \seisc. The emission line profiles
have been generated by collapsing 11 pixels of the spectra in the
direction of the resolution at the central position of the emission
lines in the rest frame, $\lambda$\,6563\,\AA, and are plotted
as a dashed line. Continuum profiles were generated by collapsing
11 resolution pixels centred at 30\,\AA\ to the red for each region and are
plotted as a dashed-dotted line. The difference between 
the two, shown as a solid line, corresponds to the pure emission. From Fig.\
\ref{profiles} it is clear that the continuum emission is not very strong,
specially in the weaker knots. The three regions are labeled in the figure.
There is a weak pure
emission knot located between Knot A and C, which does not have enough
S/N to derive the physical conditions of the gas.


\section{Results}
\label{results}


\subsection{Line intensities and reddening correction}
\label{hii:linered}

\begin{figure*}
\includegraphics[width=.48\textwidth,height=.31\textwidth,angle=0]{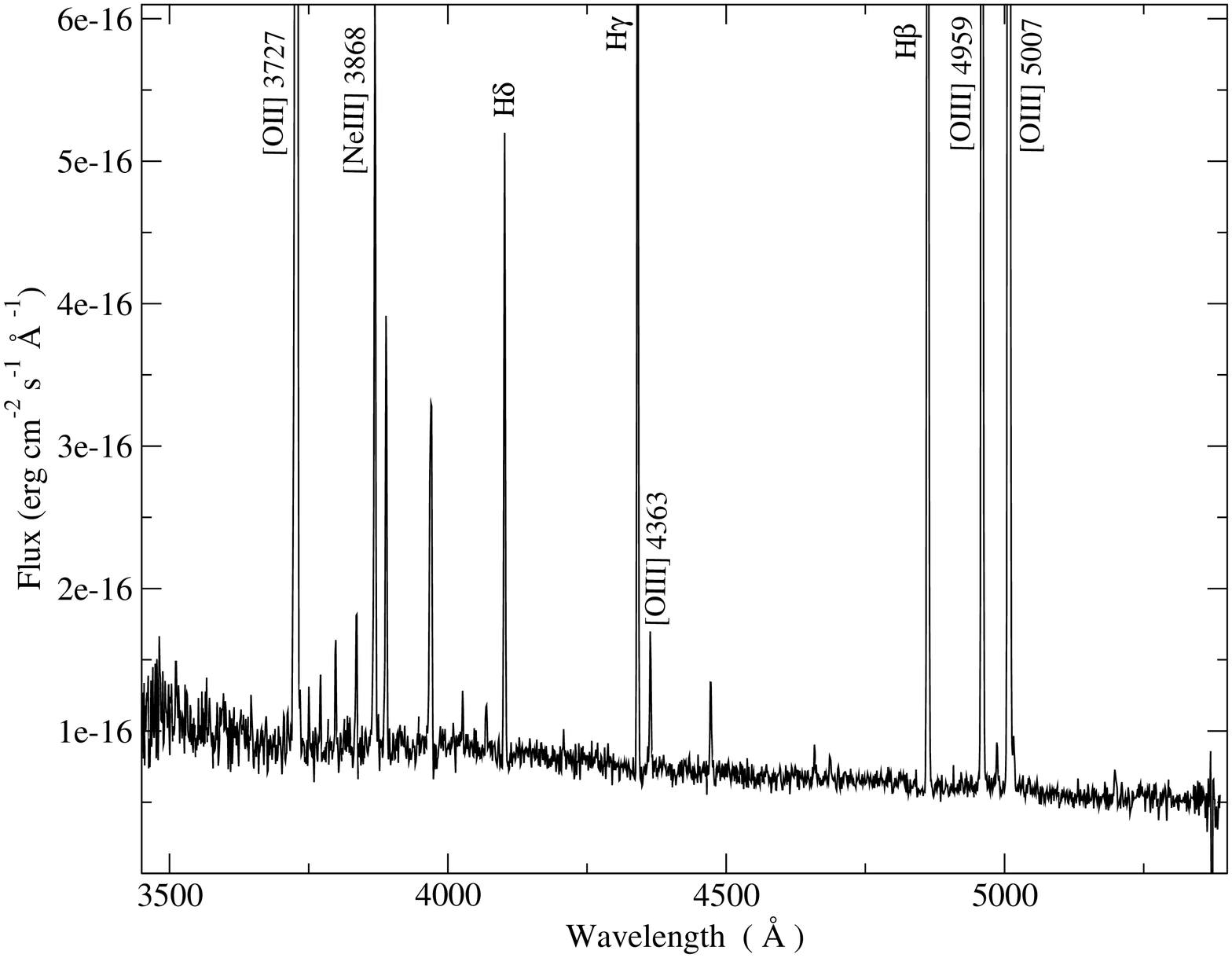}\hspace{0.5cm}
\includegraphics[width=.48\textwidth,height=.31\textwidth,angle=0]{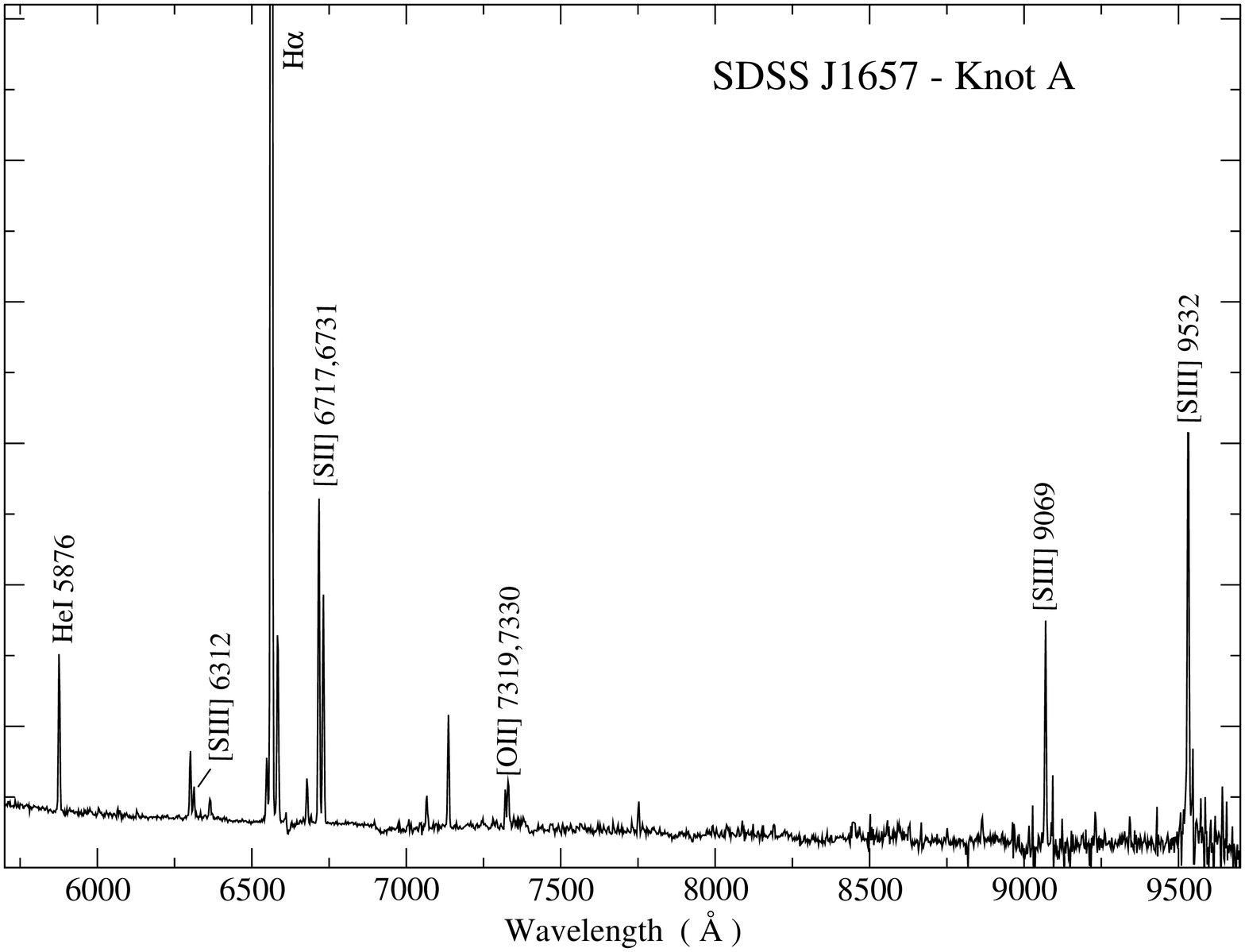}\\
\vspace{0.2cm}
\includegraphics[width=.48\textwidth,height=.31\textwidth,angle=0]{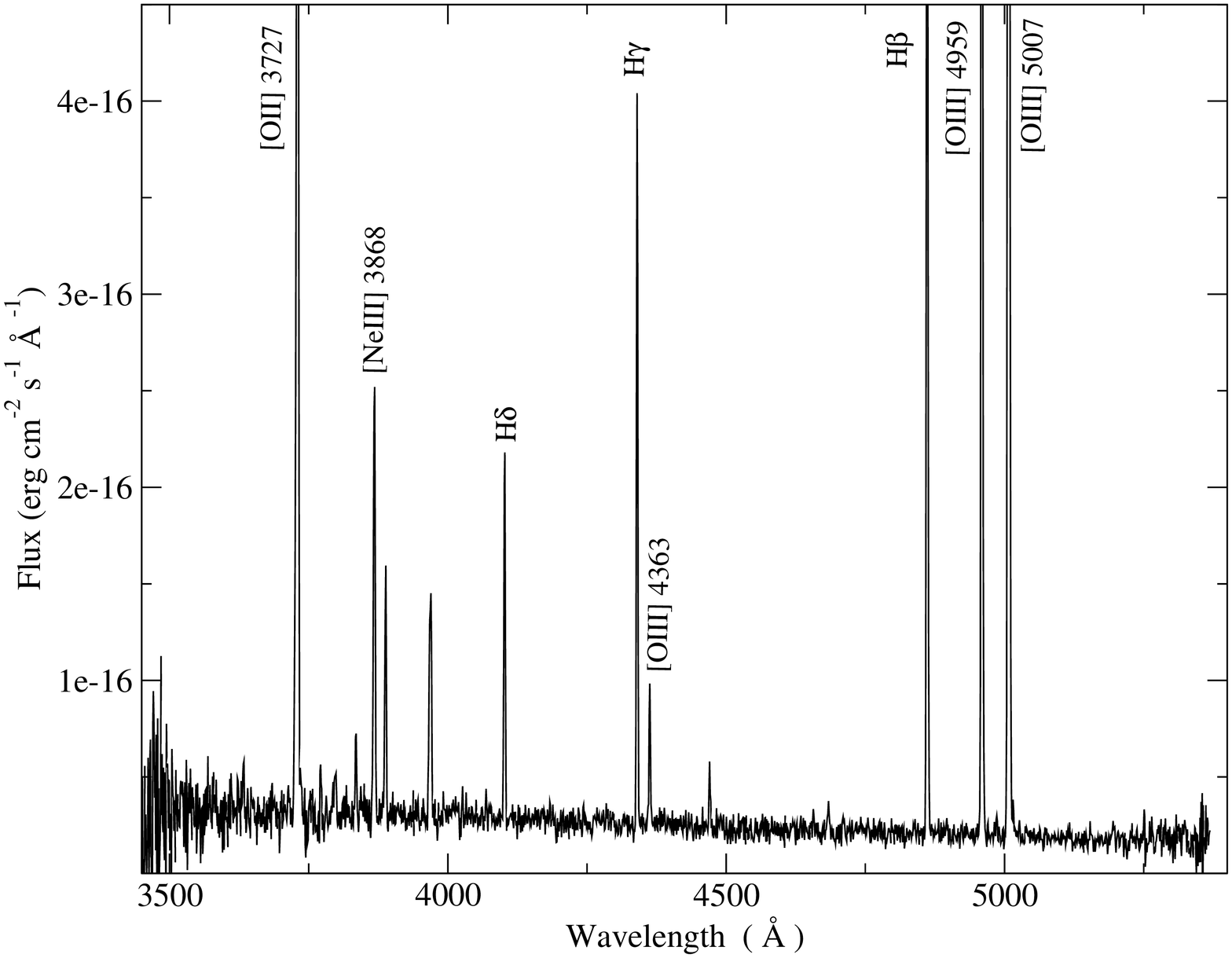}\hspace{0.5cm}
\includegraphics[width=.48\textwidth,height=.31\textwidth,angle=0]{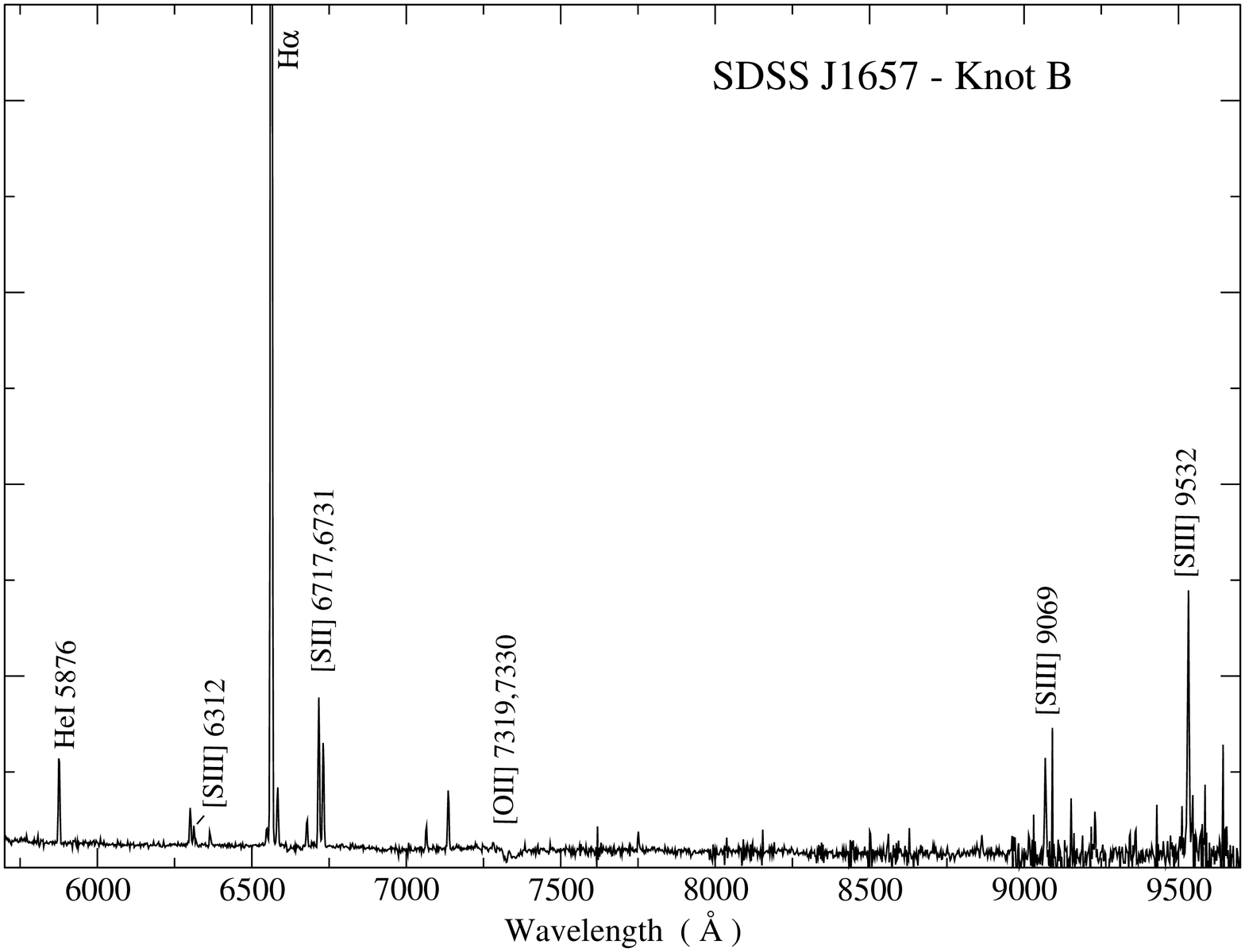}\\
\vspace{0.2cm}
\includegraphics[width=.48\textwidth,height=.31\textwidth,angle=0]{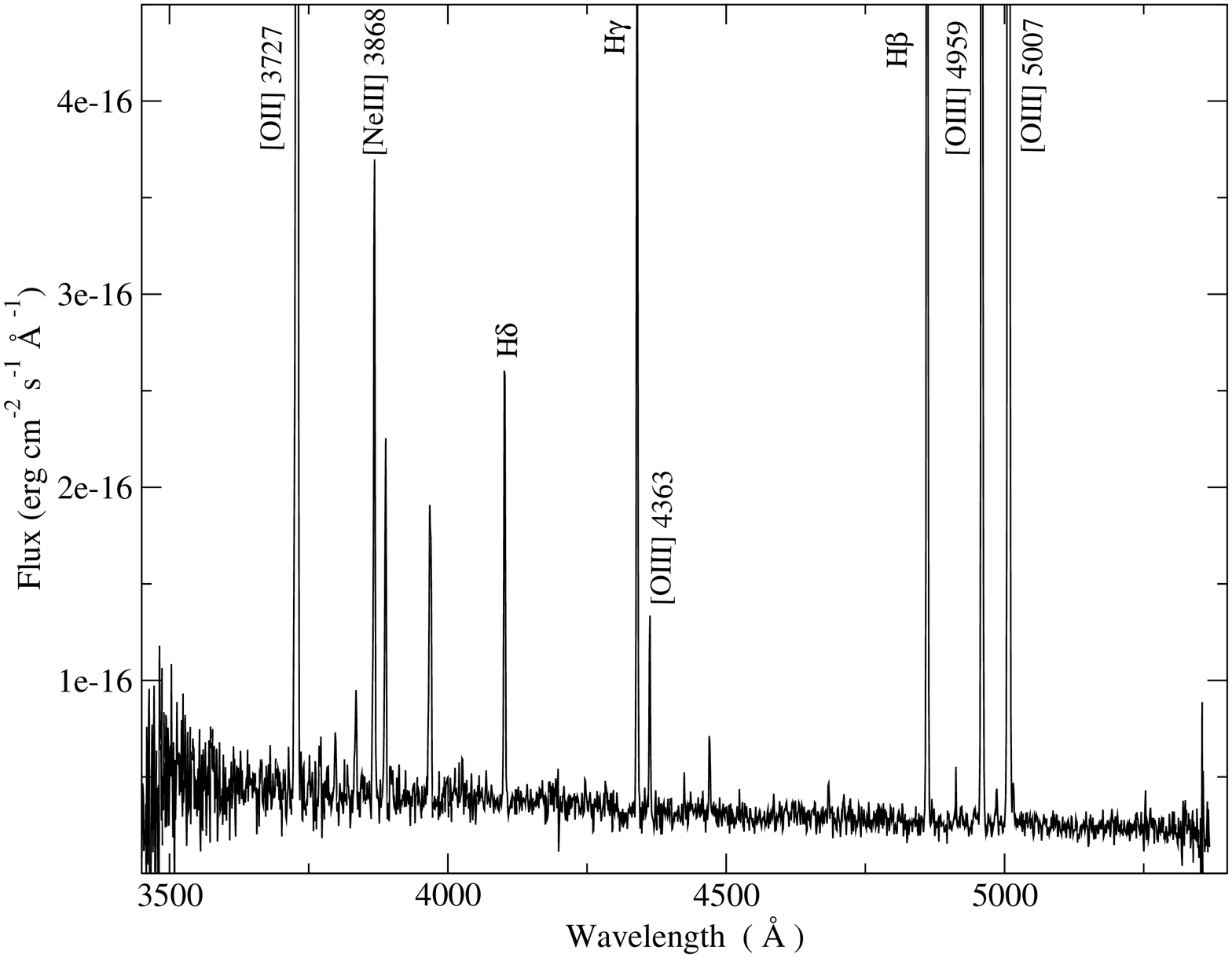}\hspace{0.5cm}
\includegraphics[width=.48\textwidth,height=.31\textwidth,angle=0]{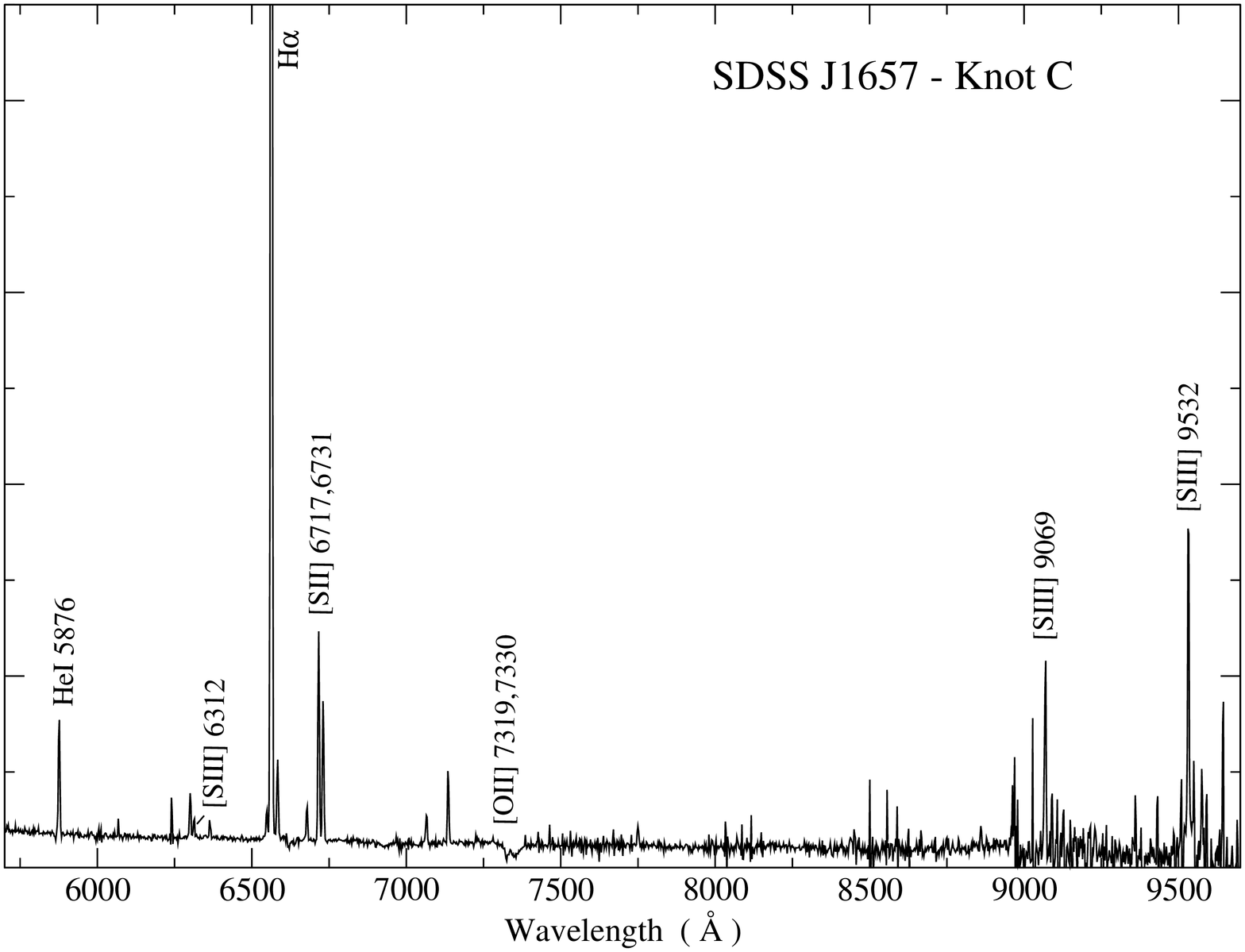}\\
\caption{Blue and red  spectra of Knots A, B and C of \seisc (upper, middle
  and lower panel, respectively) in the rest frame. The flux scales are the
  same in both spectral ranges.} 
\label{specseis}
\end{figure*}

The spectra of the three knots of \seisc (labelled from A to C) with some of
the relevant identified emission lines are shown in Fig.\ \ref{specseis}. The
spectrum of each observed knot is split into two panels. Knot A 
corresponds to the one analysed in Paper II. 

The  emission line fluxes were measured using the \texttt{splot} task in {\sc
  iraf} following 
the procedure described in Paper I. Following \cite{Perez-Montero+03}, the
statistical errors associated with the observed emission fluxes have been
calculated using the expression  \[ \sigma_{l}\,=\,\sigma_{c}N^{1/2}[1 +
  EW/(N\Delta)]^{1/2} \] \noindent where $\sigma_{l}$ is the error in the
observed line flux, $\sigma_{c}$ represents the standard deviation in a box
near the measured emission line and stands for the error in the continuum
placement, N is the number of pixels used in the measurement of the line flux,
EW is the line equivalent width, and $\Delta$ is the wavelength dispersion in
\AA\ per pixel \citep{1994ApJ...437..239G}. There are several emission lines
affected by cosmetic faults or charge transfer in the CCD, internal
reflections in the spectrograph, telluric emission lines or atmospheric
absorption lines. These cause the errors to increase, and, in some cases, they
are impossible to quantify, in which case they were ignored and excluded from any
subsequent analysis. 

Some observed lines (e.g., [Cl{\sc
iii}]\,$\lambda\lambda$\,5517,5537, several carbon recombination lines, Balmer
or Paschen lines) were impossible to 
measure due to  low signal to noise. This is also the case for the Balmer and
Paschen jump, that could not be measured because of the difficulty to fit the
continuum at both sides of the discontinuity with an acceptable precision. 

An underlying stellar population is easily appreciable by the presence of
absorption features that depress the Balmer and Paschen emission lines. A
pseudo-continuum has been defined at the base of the hydrogen emission lines
to measure the line intensities and minimize the errors introduced by the
underlying population (see Paper I). The presence of the wings of the
absorption lines imply that, even though we have used a pseudo-continuum,
there is still an absorbed fraction of the emitted flux that we are not able
to measure accurately \citep[see discussion
in][]{1988MNRAS.231...57D}. This fraction is not the same for all lines, nor
are the ratios between the absorbed fractions and the emission. In Paper I we
estimated that the difference between the measurements obtained using the
defined pseudo-continuum or a multi-Gaussian fit to the
absorption and emission components is, for
all the Balmer lines, within the errors. 
This is also the case for the objects studied here. At any
rate, for the Balmer and Paschen emission lines we have doubled the derived
error, $\sigma_{l}$, as a conservative approach to include the uncertainties
introduced by the presence of the underlying stellar population.  

The absorption features of the underlying stellar population may also affect
the helium emission lines to some extent. However, these absorption lines are
narrower than those of hydrogen \cite[see, for
  example,][]{2005MNRAS.357..945G}. Therefore it is difficult to set adequate
pseudo-continua at both sides of the lines to measure their fluxes.

We also applied the STARLIGHT code\footnote{The STARLIGHT project is supported by
the Brazilian agencies CNPq, CAPES and  FAPESP and by the France-Brazil
CAPES/Cofecub program.} \citep{2005MNRAS.358..363C} to each region to separate
the emission spectra from the underlying stellar absorptions, but for 
the strongest emission lines the difference between the measurements made
after the subtraction of the STARLIGHT fit and the ones made using the
pseudo-continuum is well below the observational errors. For a detailed
discussion on the differences in the emission line measurements see
\cite{Perez-Montero+10}. 


\begin{table*}
\centering
\caption[]
{Relative reddening corrected line intensities [$F(H\beta)$=$I(H\beta)$=10000] for the three 
star-forming knots.}
\label{ratiostot}
\begin{tabular}{@{}lcccccccccc@{}}
\hline
 &  & \multicolumn{3}{c}{\hspace{-2em} Knot A} & \multicolumn{3}{c}{\hspace{-2em} Knot B} & 
 	\multicolumn{3}{c}{ Knot C} \\
\cline{3-5} \cline{6-8} \cline{9-11}
\multicolumn{1}{c}{$\lambda$ ({\AA})} & f($\lambda$)
 & EW  & $I(\lambda)$ & Error  & EW  & $I(\lambda)$ & Error  & EW  & $I(\lambda)$ & Error  \\
&  \multicolumn{1}{c}{} & \multicolumn{1}{c}{(\AA)} & \multicolumn{1}{c}{} & \multicolumn{1}{c}{\hspace{-1em}(\%)} & 
\multicolumn{1}{c}{(\AA)} & \multicolumn{1}{c}{} & \multicolumn{1}{c}{\hspace{-1em}(\%)} & \multicolumn{1}{c}{(\AA)} & 
\multicolumn{1}{c}{} & \multicolumn{1}{c}{(\%)}  \\
\hline
 
 3727 [\OII]$^a$      &   0.271   &  120.1 & 18832\,$\pm$\,230 &  1.2 &  107.8    & 13266\,$\pm$\,420 &  3.2     &  112.2    & 14809\,$\pm$\,282 &  1.9     \\
 3750 H12             &   0.266   &    1.9 &   232\,$\pm$\, 47 & 20.2 &    ---    &          ---    &   ---    &    ---    &    ---          &   ---    \\
 3770 H11             &   0.261   &    2.3 &   293\,$\pm$\, 40 & 13.8 &    9.7    &   617\,$\pm$\,146 & 23.7     &    ---    &    ---          &   ---    \\
 3798 H10             &   0.254   &    4.1 &   500\,$\pm$\, 68 & 13.5 &    ---    &          ---    &   ---    &    4.9    &   465\,$\pm$\,141 & 30.4     \\
 3835 H9              &   0.246   &    7.1 &   780\,$\pm$\, 93 & 11.9 &    8.9    &   783\,$\pm$\,180 & 23.0     &    7.5    &   761\,$\pm$\,140 & 18.4     \\
 3868 [\NeIII]        &   0.238   &   23.1 &  3262\,$\pm$\,132 &  4.0 &   30.7    &  3663\,$\pm$\,132 &  3.6     &   28.6    &  3533\,$\pm$\,118 &  3.3     \\
 3889 \HeI+H8         &   0.233   &   14.0 &  1826\,$\pm$\, 95 &  5.2 &   22.3    &  1954\,$\pm$\,208 & 10.7     &   26.2    &  2192\,$\pm$\,267 & 12.2     \\
 3968 [\NeIII]+H7     &   0.216   &   22.4 &  2456\,$\pm$\,121 &  4.9 &   30.1    &  2705\,$\pm$\,221 &  8.2     &   29.9    &  2699\,$\pm$\,237 &  8.8     \\
 4026 [\NII]+\HeI     &   0.203   &    1.1 &   155\,$\pm$\, 16 & 10.4 &    4.3    &   327\,$\pm$\,105 & 32.1     &    ---    &    ---          &   ---    \\
 4068 [\SII]          &   0.195   &    1.4 &   198\,$\pm$\, 15 &  7.4 &    1.4    &   153\,$\pm$\, 36 & 23.7     &    1.2    &   130\,$\pm$\, 35 & 26.9     \\
 4102 H$\delta$       &   0.188   &   20.9 &  2432\,$\pm$\, 65 &  2.7 &   30.2    &  2667\,$\pm$\,163 &  6.1     &   29.2    &  2619\,$\pm$\,173 &  6.6     \\
 4340 H$\gamma$       &   0.142   &   43.1 &  4417\,$\pm$\, 97 &  2.2 &   76.5    &  4813\,$\pm$\,219 &  4.5     &   68.7    &  4712\,$\pm$\,140 &  3.0     \\
 4363 [\OIII]         &   0.138   &    4.5 &   524\,$\pm$\, 24 &  4.6 &    8.3    &   846\,$\pm$\, 67 &  7.9     &    8.9    &   838\,$\pm$\, 59 &  7.0     \\
 4471 \HeI            &   0.106   &    4.2 &   443\,$\pm$\, 33 &  7.4 &    4.7    &   405\,$\pm$\, 44 & 11.0     &    5.3    &   445\,$\pm$\, 41 &  9.3     \\
 4658 [\FeIII]        &   0.053   &    1.0 &   107\,$\pm$\, 16 & 14.9 &    ---    &          ---    &   ---    &    ---    &    ---          &   ---    \\
 4686 \HeII           &   0.045   &    1.2 &   126\,$\pm$\, 14 & 11.0 &    4.3    &   313\,$\pm$\, 68 & 21.9     &    2.1    &   170\,$\pm$\, 24 & 14.2     \\
 4861 H$\beta$        &   0.000   &  117.8 & 10000\,$\pm$\, 79 &  0.8 &  153.3    & 10000\,$\pm$\,178 &  1.8     &  167.2    & 10000\,$\pm$\,128 &  1.3     \\
 4921 \HeI            &  -0.014   &    0.8 &    75\,$\pm$\, 14 & 18.6 &    ---    &          ---    &   ---    &    ---    &    ---          &   ---    \\
 4959 [\OIII]         &  -0.024   &  152.5 & 14333\,$\pm$\,127 &  0.9 &  218.5    & 16118\,$\pm$\,129 &  0.8     &  199.2    & 14940\,$\pm$\,114 &  0.8     \\
 4986 [\FeIII]$^b$    &  -0.030   &    1.4 &   135\,$\pm$\, 28 & 20.5 &    2.4    &   163\,$\pm$\, 57 & 35.1     &    3.2    &   215\,$\pm$\, 39 & 18.3     \\
 5007 [\OIII]         &  -0.035   &  455.1 & 43082\,$\pm$\,240 &  0.6 &  705.5    & 48653\,$\pm$\,256 &  0.5     &  613.2    & 44727\,$\pm$\,129 &  0.3     \\
 5015 \HeI            &  -0.037   &    2.4 &   222\,$\pm$\, 23 & 10.1 &    3.2    &   205\,$\pm$\, 35 & 17.3     &    3.2    &   220\,$\pm$\, 28 & 12.6     \\
 5199 [\NI]           &  -0.078   &    2.0 &   157\,$\pm$\, 26 & 16.4 &    ---    &          ---    &   ---    &    ---    &    ---          &   ---    \\
 5876 \HeI             &  -0.209   &   18.9 &  1116\,$\pm$\, 44 &  3.9 &   23.4    &   991\,$\pm$\, 29 &  2.9     &   29.0    &  1149\,$\pm$\, 47 &  4.1     \\
 6300 [\OI]            &  -0.276   &    8.1 &   438\,$\pm$\, 16 &  3.7 &    9.6    &   393\,$\pm$\, 26 &  6.7     &    8.6    &   366\,$\pm$\, 14 &  3.9     \\
 6312 [\SIII]          &  -0.278   &    3.7 &   201\,$\pm$\,  9 &  4.3 &    4.3    &   175\,$\pm$\, 10 &  5.8     &    3.5    &   148\,$\pm$\, 10 &  6.4     \\
 6364 [\OI]            &  -0.285   &    2.8 &   152\,$\pm$\, 18 & 11.8 &    3.6    &   141\,$\pm$\, 20 & 14.4     &    3.0    &   124\,$\pm$\, 12 & 10.0     \\
 6548 [\NII]           &  -0.311   &    9.4 &   464\,$\pm$\, 23 &  5.0 &    5.5    &   216\,$\pm$\, 23 & 10.7     &    6.7    &   266\,$\pm$\, 14 &  5.4     \\
 6563 H$\alpha$       &  -0.313   &  571.3 & 27772\,$\pm$\,153 &  0.5 &  772.5    & 28159\,$\pm$\,105 &  0.4     &  730.5    & 27919\,$\pm$\,133 &  0.5     \\
 6584 [\NII]           &  -0.316   &   28.8 &  1428\,$\pm$\, 47 &  3.3 &   16.2    &   632\,$\pm$\, 50 &  8.0     &   17.0    &   680\,$\pm$\, 29 &  4.3     \\
 6678 \HeI             &  -0.329   &    6.7 &   315\,$\pm$\, 18 &  5.7 &    7.3    &   272\,$\pm$\, 11 &  4.1     &    7.9    &   296\,$\pm$\, 20 &  6.9     \\
 6717 [\SII]           &  -0.334   &   47.4 &  2207\,$\pm$\, 57 &  2.6 &   39.4    &  1489\,$\pm$\, 35 &  2.3     &   45.1    &  1699\,$\pm$\, 56 &  3.3     \\
 6731 [\SII]           &  -0.336   &   32.2 &  1598\,$\pm$\, 43 &  2.7 &   28.1    &  1060\,$\pm$\, 25 &  2.3     &   30.7    &  1154\,$\pm$\, 40 &  3.4     \\
 7065 \HeI             &  -0.377   &    5.6 &   235\,$\pm$\, 10 &  4.4 &    7.5    &   226\,$\pm$\, 12 &  5.1     &    8.8    &   279\,$\pm$\, 26 &  9.3     \\
 7136 [\ArIII]         &  -0.385   &   16.4 &   717\,$\pm$\, 26 &  3.6 &   18.4    &   584\,$\pm$\, 19 &  3.3     &   17.6    &   581\,$\pm$\, 17 &  2.9     \\
 7281 \HeI$^c$         &  -0.402   &    0.9 &    41\,$\pm$\,  7 & 18.2 &    3.0    &    90\,$\pm$\, 14 & 15.9     &    1.4    &    46\,$\pm$\, 13 & 28.6     \\
 7319 [\OII]$^d$       &  -0.406   &   12.3 &   302\,$\pm$\, 17 &  5.6 &    4.8    &   165\,$\pm$\, 15 &  9.2     &    6.1    &   196\,$\pm$\, 18 &  9.2     \\
 7330 [\OII]$^e$       &  -0.407   &    8.8 &   211\,$\pm$\, 14 &  6.5 &    7.2    &   251\,$\pm$\, 21 &  8.4     &    7.9    &   254\,$\pm$\, 26 & 10.1     \\
 7751 [\ArIII]         &  -0.451   &    4.6 &   177\,$\pm$\, 22 & 12.3 &    5.5    &   163\,$\pm$\, 13 &  8.0     &    5.8    &   170\,$\pm$\, 17 &  9.9     \\
 8665 P13             &  -0.531   &    7.7 &   144\,$\pm$\, 53 & 37.1 &    ---    &          ---    &   ---    &    8.9    &   133\,$\pm$\, 38 & 28.3     \\
 8751 P12             &  -0.537   &    4.2 &   101\,$\pm$\, 29 & 28.2 &    ---    &          ---    &   ---    &    ---    &    ---          &   ---    \\
 8865 P11             &  -0.546   &    8.5 &   211\,$\pm$\, 34 & 16.3 &   17.9    &   277\,$\pm$\, 73 & 26.2     &   11.3    &   228\,$\pm$\, 46 & 20.2     \\
 9014 P10             &  -0.557   &   15.4 &   167\,$\pm$\, 36 & 21.5 &    ---    &          ---    &   ---    &    ---    &    ---          &   ---    \\
 9069 [\SIII]          &  -0.561   &   59.2 &  1400\,$\pm$\, 99 &  7.1 &   68.2    &   984\,$\pm$\,122 & 12.4     &   81.9    &  1577\,$\pm$\,134 &  8.5     \\
 9229 P9              &  -0.572   &   16.9 &   263\,$\pm$\, 47 & 18.0 &   26.1    &   364\,$\pm$\,129 & 35.6     &    ---    &      ---        &   ---    \\
 9532 [\SIII]          &  -0.592   &  157.3 &  3674\,$\pm$\,257 &  7.0 &   85.1    &  2700\,$\pm$\,152 &  5.6     &  238.7    &  2915\,$\pm$\,162 &  5.6     \\
\hline
\multicolumn{2}{l}{I(H$\beta$)(erg\,sec$^{-1}$\,cm$^{-2}$)} & \multicolumn{3}{c}{\hspace{-1em}6.3\,$\times$\,10$^{-15}$}  & 
\multicolumn{3}{c}{\hspace{-1em}2.6\,$\times$\,10$^{-15}$}  & \multicolumn{3}{c}{3.4\,$\times$\,10$^{-15}$}\\
\multicolumn{2}{l}{c(H$\beta$)}  & \multicolumn{3}{c}{\hspace{-2em} 0.05\,$\pm$\,0.01 }    & 
\multicolumn{3}{c}{\hspace{-2em} 0.15\,$\pm$\,0.02 }    & \multicolumn{3}{c}{ 0.13\,$\pm$\,0.02 }\\
\hline
\noalign {\noindent   
$^a$\,[O{\sc ii}]\,$\lambda\lambda$\,3726\,+\,3729;
$^b$\,[Fe{\sc iii}]\,$\lambda\lambda$\,4986\,+\,4987;
$^c$\,possibly blend with an unknown line; 
$^d$\,[O{\sc ii}]\,$\lambda\lambda$\,7318\,+\,7320; 
$^e$\,[O{\sc ii}]\,$\lambda\lambda$\,7330\,+\,7331.}

\end{tabular}
\end{table*}


The reddening coefficients $c$(H$\beta$) were calculated from the measured
Balmer decrements, $F(\lambda)$/$F(H\beta)$. We adopted the galactic
extinction law of \cite{Miller+72} with $R_{\rm v}$=3.2. A least square fit of
the measured decrements to the theoretical ones, $(F(\lambda)$/$F(H\beta))_0$,
computed based on the data by \cite{Storey+95}, was performed that provides
the value of $c$(H$\beta$). The theoretical Balmer decrements depend on
electron density and temperature. We used an iterative method to estimate
them, taking as starting values those derived from the measured [S{\sc ii}]
$\lambda\lambda$ 6717,6731\,\AA\  and  [O{\sc iii}] $\lambda\lambda$ 4363,
4959, 5007\,\AA\ line fluxes. Due to the large error introduced by the
presence of the underlying stellar population, only the strongest Balmer
emission lines (H$\alpha$, H$\beta$, H$\gamma$ and H$\delta$) were used. 

{For the easiness of comparison}, we have included in the following sections 
the results presented in Paper II for knot A.
Table \ref{ratiostot}  lists the reddening corrected emission lines for
each knot, together with the reddening constant and its error taken
as the uncertainties of the least square fit and the reddening corrected
H$\beta$ intensity. Column 1 shows the wavelength and the name
of the measured lines. The adopted reddening curve, $f(\lambda)$, normalized
to H$\beta$, is given in column 2. The errors in the emission lines were
obtained by propagating in quadrature the observational errors in the emission
line fluxes and the reddening constant uncertainties. We have not taken into
account errors in the theoretical intensities since they are much lower than
the observational ones.

\subsection{Physical conditions of the gas}

The physical conditions of the ionized gas, including electron temperatures
(T$_e$) and electron density (N$_e$\,$\approx$\,n([\SII])), have been derived
from the emission line data using the same procedures as in Paper II, based on
the five-level statistical equilibrium atom approximation in the task
\texttt{temden}, of the software package IRAF
\citep{1987JRASC..81..195D,1995PASP..107..896S}. As usual, we have taken as 
sources of error the uncertainties associated with the measurement of the
emission-line fluxes and the reddening correction, and we have propagated them
through our calculations.  

For all three knots we have derived the electron temperatures of
[O{\sc ii}], [O{\sc iii}], [S{\sc ii}] and [S{\sc iii}]. The [O{\sc
    ii}]\,$\lambda\lambda$\,7319,7330\,\AA\ lines have a contribution by
direct recombination which increases with temperature. Using the calculated
[O{\sc iii}] electron temperatures, we have estimated these contributions to
be less than 4\,\% in all cases and therefore we have not corrected for this
effect. The expression for the  correction of direct recombination, however,
is valid only in the range of temperatures between 5000 and 10000\,K. The
temperatures found are slightly over that range. At any rate, the relative
contribution of recombination to collisional intensities decreases rapidly
with increasing temperature, therefore for the high T$_e$  values found in our
objects this contribution is expected to be small.

The derived electron densities and temperatures for the three star-forming
regions are given in Table \ref{jhtemden} along with their corresponding 
errors.

\begin{table}
\centering 
\caption{Electron densities and temperatures. Densities 
in cm$^{-3}$ and temperatures in 10$^{4}$ K.}
\begin{tabular}{@{}l@{\hspace{0.25cm}}c@{\hspace{0.25cm}}c@{\hspace{0.25cm}}c@{\hspace{0.25cm}}c@{\hspace{0.25cm}}c@{}}
\hline
& n([S{\sc ii}]) & T$_e$([O{\sc iii}]) &  T$_e$([O{\sc ii}]) & T$_e$([S{\sc iii}]) & T$_e$([S{\sc ii}]) \\
\hline
Knot A &  30:  & 1.23$\pm$0.02 & 1.33$\pm$0.07 & 1.45$\pm$0.08 & 0.88$\pm$0.05  \\
Knot B &  10:  & 1.43$\pm$0.05 & 1.52$\pm$0.12 & 1.64$\pm$0.11 & 1.00$\pm$0.17  \\
Knot C &  10:  & 1.48$\pm$0.05 & 1.50$\pm$0.13 & 1.29$\pm$0.07 & 0.83$\pm$0.15  \\
\hline
\end{tabular}
\label{jhtemden}
\end{table}

\begin{table}
\centering
\caption[Ionic and total helium abundances]
{Ionic and total helium abundance.}
\begin{tabular}{@{}l@{\hspace{0.2cm}}c@{\hspace{0.2cm}}c@{\hspace{0.2cm}}c@{}}
\hline
    &   Knot A &     Knot B &    Knot C  \\
\hline
He$^{+}$/H$^{+}$($\lambda$4471)  &  0.093$\pm$0.007   & 0.085$\pm$0.009   & 0.094$\pm$0.009  \\  
He$^{+}$/H$^{+}$($\lambda$5876)  &  0.085$\pm$0.003   & 0.079$\pm$0.002   & 0.093$\pm$0.003  \\
He$^{+}$/H$^{+}$($\lambda$6678)  &  0.086$\pm$0.005   & 0.077$\pm$0.003   & 0.084$\pm$0.005  \\   
He$^{+}$/H$^{+}$($\lambda$7065)  &  0.093$\pm$0.005   & 0.086$\pm$0.006   & 0.105$\pm$0.010  \\  
He$^{+}$/H$^{+}$(Adop.)        &  0.087$\pm$0.005   & 0.080$\pm$0.005   & 0.092$\pm$0.009  \\  
He$^{2+}$/H$^{+}$($\lambda$4686) &  0.0011$\pm$0.0001 & 0.0028$\pm$0.0006 & 0.0015$\pm$0.0002 \\ 
(He/H)                            &  0.088$\pm$0.008   & 0.080$\pm$0.008   & 0.092$\pm$0.009    \\ 
\hline
\end{tabular}
\label{jhheab}
\end{table}

\subsection{Chemical abundance derivation}

We have derived the ionic chemical abundances of the different species using
the strongest available emission lines detected in the analyzed spectra and
the task \texttt{ionic} of the STSDAS package in IRAF, as described in Paper
II. 

The total abundances have been calculated by taking into account, when
required, the unseen ionization stages of each element, using the appropriate
ionization correction factor (ICF) for each species, 
X/H\,=\,ICF(X$^{+i}$)\,X$^{+i}$/H${^+}$ as detailed in what follows.

\subsubsection{Helium}

We have used the well detected  He{\sc i}\,$\lambda\lambda$\,4471,
5876, 6678 and 7065\,\AA\ lines, to calculate the abundances of once ionized
helium. For the three knots also the He{\sc ii}\,$\lambda$\,4686\,\AA\ line
was measured allowing the calculation of twice ionized He. The He lines arise
mainly from pure recombination, although they could have some contribution
from collisional excitation and be affected by self-absorption \citep[see][for
a complete treatment of these effects]{2001NewA....6..119O,
  2004ApJ...617...29O}. We have taken the electron temperature of [O{\sc iii}]
as representative of the zone where the He emission arises since at any rate
ratios of recombination lines are weakly sensitive to electron temperature. We
have used the equations given by Olive \& Skillman  to derive the
He$^{+}$/H$^{+}$ value, using the theoretical emissivities scaled to H$\beta$
from \cite{1999ApJ...514..307B} and the expressions for the collisional
correction factors from \cite{1995ApJ...442..714K}. To calculate the abundance
of twice ionized helium we have used equation (9) from
\cite{Kunth+83}. We have not made any 
corrections for fluorescence since three of the used helium lines have a small
dependence with optical depth effects but the observed  objects have low
densities. We have not corrected either for the presence of an underlying stellar population. 
A summary of the equations used to calculate these ionic abundances is given in
Appendix B of \cite{RubenPhD}.
The total abundance of He has been found by adding directly the two ionic
abundances,  He/H\,=\,(He${^+}$+He$^{2+}$)/H$^+$.
The results obtained for each line and the total He abundances, along with
their corresponding errors are presented in Table \ref{jhheab}. Also given in
the table is the adopted value for He$^{+}$/H$^{+}$  as the average, weighted
by the errors, of the abundances derived from each He{\sc i} emission line . 

\subsubsection{Ionic and total chemical abundances from forbidden lines}

The oxygen ionic abundance ratios, O$^{+}$/H$^{+}$ and O$^{2+}$/H$^{+}$, have
been derived from the [O{\sc ii}]\,$\lambda\lambda$\,3727,29\,\AA\ and [O{\sc
iii}]\,$\lambda\lambda$\,4959, 5007\,\AA\ lines respectively using for each
ion its corresponding temperature. At the temperatures derived  here, 
most of the oxygen is in the form of O$^+$ and
O$^{2+}$, therefore the approximation O/H\,=\,(O$^+$+O$^{2+}$)/H$^+$ 
is a valid one. 

S$^+$/H$^{+}$ and S$^{2+}$/H$^{+}$,
abundances have been derived using T$_e$([S{\sc ii}]) and T$_e$([S{\sc iii}])
and the fluxes of 
the [S{\sc ii}]\,$\lambda\lambda$\,6717,6731\,{\AA} and  the
near-IR [S{\sc iii}]\,$\lambda\lambda$\,9069, 9532\,\AA\  emission lines,
respectively. Unlike oxygen,  a relatively important contribution
from S$^{3+}$ may be expected for sulphur depending on the nebular
excitation. The total 
sulphur abundance has been calculated using an ICF for S$^+$+S$^{2+}$
according to the formula by \cite{Barker80}, which is based on \cite{Stasinska78}
photo-ionization models, 
with $\alpha$\,=\,2.5, which gives the best fit to the scarce
observational data on S$^{3+}$ abundances \citep{Perez-Montero+06}. 
Taking this ICF as
a function of the ratio O$^{2+}$/O instead of O$^+$/O reduces the propagated error
for this quantity.

The ionic abundance of nitrogen, N$^{+}$/H$^{+}$ has been derived from the 
intensities of the [NII]$\lambda\lambda$\,6548,6584\,\AA\ lines  
assuming T$_e$([N{\sc ii}])\,$\approx$\,T$_e$([O{\sc ii}]). The N/O
abundance ratio has been derived under the assumption that N/O\,=\,N$^+$/O$^+$
and  N/H was calculated as log(N/H)\,=\,log(N/O)+log(O/H).

Neon is only visible in the spectra via the [Ne{\sc iii}] emission line at
$\lambda$3868\,{\AA}, so Ne$^{2+}$ has been derived using this line. For this ion we have 
taken the electron temperature of [O{\sc iii}], as representative of the high excitation 
zone \citep[T$_e$({[}Ne{\sc iii}{]})\,$\approx$\,T$_e$({[}O{\sc iii}{]});][]{Peimbert+69}.
Classically, the total abundance of neon has been calculated assuming that
Ne/O\,$\approx$\,Ne$^{2+}$/O$^{2+}$. 
\cite{Izotov+04} point out that this assumption can lead to an overestimate of 
Ne/H in objects with low excitation, where the charge transfer between O$^{2+}$
and H$^0$ becomes important. Thus, we have used the expression of
this ICF given by \citep{Perez-Montero+07}.
It is interesting to note, however, that given the high excitation of the
observed objects there is no significant difference between the total neon
abundance derived using this ICF and those estimated using the classic
approximation.

The only available emission lines of argon in the optical spectra of ionized
regions correspond to Ar$^{2+}$ and Ar$^{3+}$. The abundance of Ar$^{2+}$ has
been calculated from the measured [Ar{\sc iii}]\,$\lambda$\,7136\,\AA\ line
assuming that T$_e$([Ar{\sc iii}])\,$\approx$\,T$_e$([S{\sc iii}])
\citep{Garnett92}. [Ar{\sc iv}] was not detected in the spectra. 
The total abundance of Ar was hence calculated  using the ICF(Ar$^{2+}$)
derived from photo-ionization models by \cite{Perez-Montero+07}.

The ionic abundances with respect to ionized hydrogen of the  elements heavier than helium, ICFs,
total abundances and their corresponding errors are given in Table \ref{jhabicf}.  

\begin{table}
\begin{minipage}[t]{\columnwidth}
\caption[]
{Ionic chemical abundances derived from forbidden emission lines, ICFs and total 
chemical abundances for elements heavier than helium.}
\label{jhabicf}
\begin{tabular}{@{}l@{\hspace{0.25cm}}c@{\hspace{0.25cm}}c@{\hspace{0.25cm}}c@{}}
\hline
                            &      Knot A            &      Knot B            &      Knot C \\
\hline
12 + log(O$^{+}$/H$^{+}$)    &       7.37\,$\pm$\,      0.09&       7.03\,$\pm$\,      0.12&       7.10\,$\pm$\,      0.12\\
12 + log(O$^{2+}$/H$^{+}$)   &       7.87\,$\pm$\,      0.02&       7.74\,$\pm$\,      0.04&       7.67\,$\pm$\,      0.04\\
\bf{12 + log(O/H)}          &       7.99\,$\pm$\,      0.04&       7.82\,$\pm$\,      0.05&       7.78\,$\pm$\,      0.06\\[2pt]
ICF(O$^+$+O$^{2+}$)$^*$     &       1.07   &   1.06   &    1.06  \\
\bf{12 + log(O/H)}$^*$      &       8.02\,$\pm$\,      0.04&       7.84\,$\pm$\,      0.05&       7.80\,$\pm$\,      0.06\\[2pt]
\hline
12 + log(S$^{+}$/H$^{+}$)    &       6.07\,$\pm$\,      0.08&       5.76\,$\pm$\,      0.19&       6.02\,$\pm$\,      0.24\\
12 + log(S$^{2+}$/H$^{+}$)   &       6.00\,$\pm$\,      0.06&       5.79\,$\pm$\,      0.07&       6.02\,$\pm$\,      0.07\\
ICF(S$^{+}$ + S$^{2+}$)      &       1.32\,$\pm$\,      0.05&       1.51\,$\pm$\,      0.10&       1.38\,$\pm$\,      0.09\\
\bf{12 + log(S/H)}           &       6.46\,$\pm$\,      0.07&       6.25\,$\pm$\,      0.13&       6.46\,$\pm$\,      0.16\\
log(S/O)                     &      -1.53\,$\pm$\,      0.08&      -1.57\,$\pm$\,      0.14&      -1.32\,$\pm$\,      0.17\\[2pt]
ICF(S$^{+}$ + S$^{2+}$)$^*$   &       1.07      &       1.18     &       1.14\\
\bf{12 + log(S/H)}$^*$       &       6.37\,$\pm$\,      0.07&       6.15\,$\pm$\,      0.13&       6.18\,$\pm$\,      0.15\\
log(S/O)$^*$            &      -1.65\,$\pm$\,      0.08&      -1.69\,$\pm$\,      0.14&      -1.42\,$\pm$\,      0.16\\[2pt]
\hline
12 + log(N$^{+}$/H$^{+}$)    &       6.15\,$\pm$\,      0.06&       5.68\,$\pm$\,      0.10&       5.74\,$\pm$\,      0.09\\
\bf{12 + log(N/H)}           &       6.76\,$\pm$\,      0.28&       6.47\,$\pm$\,      0.39&       6.42\,$\pm$\,      0.39\\
log(N/O)                     &      -1.23\,$\pm$\,      0.11&      -1.35\,$\pm$\,      0.16&      -1.36\,$\pm$\,      0.15\\[2pt]
ICF(N$^{+}$)$^*$              &       3.39   &       4.90   &       4.27 \\
\bf{12 + log(N/H)}$^*$        &       6.68\,$\pm$\,      0.06&       6.37\,$\pm$\,      0.10&       6.37\,$\pm$\,      0.09\\
log(N/O)$^*$                  &      -1.34\,$\pm$\,      0.07&      -1.47\,$\pm$\,      0.11&      -1.43\,$\pm$\,      0.11\\[2pt]
\hline
12 + log(Ne$^{2+}$/H$^{+}$)  &       7.22\,$\pm$\,      0.04&       7.06\,$\pm$\,      0.06&       7.00\,$\pm$\,      0.05\\
ICF(Ne)                      &       1.09\,$\pm$\,      0.00&       1.08\,$\pm$\,      0.00&       1.08\,$\pm$\,      0.01\\
\bf{12 + log(Ne/H)}          &       7.25\,$\pm$\,      0.04&       7.09\,$\pm$\,      0.06&       7.04\,$\pm$\,      0.05\\
log(Ne/O)                    &      -0.74\,$\pm$\,      0.06&      -0.73\,$\pm$\,      0.08&      -0.74\,$\pm$\,      0.08\\[2pt]
ICF(Ne$^{2+}$)$^*$            &       1.26  &       1.12   &       1.17   \\
\bf{12 + log(Ne/H)}$^*$      &       7.32\,$\pm$\,      0.04&       7.11\,$\pm$\,      0.06&       7.17\,$\pm$\,      0.05\\
log(Ne/O)$^*$                &      -0.70\,$\pm$\,      0.05&      -0.73\,$\pm$\,      0.08&      -0.73\,$\pm$\,      0.08\\[2pt]
\hline
12 + log(Ar$^{2+}$/H$^{+}$)  &       5.49\,$\pm$\,      0.06&       5.31\,$\pm$\,      0.06&       5.49\,$\pm$\,      0.06\\
ICF(Ar$^{2+}$)               &       1.13\,$\pm$\,      0.02&       1.23\,$\pm$\,      0.06&       1.16\,$\pm$\,      0.04\\
\bf{12 + log(Ar/H)}          &       5.54\,$\pm$\,      0.06&       5.39\,$\pm$\,      0.06&       5.55\,$\pm$\,      0.06\\
log(Ar/O)                    &      -2.45\,$\pm$\,      0.07&      -2.43\,$\pm$\,      0.08&      -2.22\,$\pm$\,      0.08\\[2pt]
ICF(Ar$^{2+}$)$^*$           &       1.15     &       1.17    &       1.17   \\
\bf{12 + log(Ar/H)}$^*$      &       5.55\,$\pm$\,      0.06&       5.38\,$\pm$\,      0.06&       5.56\,$\pm$\,      0.06\\
log(Ar/O)$^*$                &      -2.47\,$\pm$\,      0.07&      -2.46\,$\pm$\,      0.08&      -2.24\,$\pm$\,      0.08\\
\hline
\end{tabular}
\end{minipage}
$^*$ ICFs and total abundances from photoionization models (see text).  
\end{table}

\subsection{Photoionization models of the observed regions}
\label{photomod}

Detailed tailor-made photoionization models were produced
in order to ascertain the main properties of the ionizing stellar population
and the ionized gas. The  methodology is described in
\cite{Perez-Montero+10} who study the brightest knots of the H{\sc ii}
galaxies described in Papers I and II, including knot A in \seisc. Here
 we describe the models for knots B and C, and compare them
with the observations and with the results obtained for knot A in
\cite{Perez-Montero+10}. 

We have resorted to the photoionization code CLOUDY v. 06.02c
\citep{Ferland+98}, using the equivalent width of H$\beta$, after removing the 
underlying stellar population ({\em i.e.}
the population younger than 10 Myr), the H$\alpha$ luminosity and the 
 intensities of [O{\sc ii}]\,3727\,\AA, [O{\sc iii}]\,4363 and
5007\,\AA, [S{\sc ii}]\,6717 and 6731\,\AA, and [S{\sc iii}]\,9069 and
9532\,\AA\ relative to H$\beta$.

\begin{figure*}
\centering
\includegraphics[width=12cm]{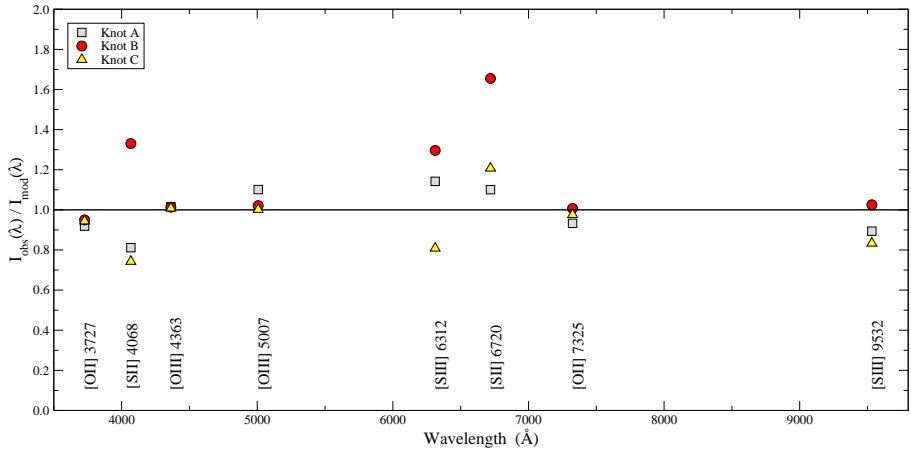}
\caption[]
{Ratio between observed and modelled intensities of the most representative
emission lines for each one of the star-formig knots.}
\label{mod_lin}
\end{figure*}

We have used for the photoionization models the same Starburst 99
stellar libraries as in the model fitting of the stellar population,
described in \S\ref{stpop}, with the metallicity closest  to the value
measured in the gas-phase, Z = 0.004 (= 1/5 Z$_{\odot}$). We assummed a constant
star formation history which, according to P\'erez-Montero et al., gives the
best agreement for the number of ionizing photons and the EW(H$\beta$)
corrected for underlying stellar population and dust absorption effects. A
thick shell geometry and a constant density of 100 particles per cm$^3$ have
been set as input conditions in all the models. To fit the observed
properties, the distance to the ionizing
source, the filling factor, the dust-to-gas ratio and the age of the stellar
cluster were left as free parameters.

One of the most
important parameters in the correct modelling of ionized gas nebulae is the
dust absorption factor, $f_d$, which gives the ratio between the number of
ionizing photons emitted by the stellar cluster and the number of ionizing
photons absorbed by the gas \citep{Perez-Montero+10}. This factor must be taken into account in 
deriving properties of the cluster from hydrogen Balmer recombination
lines. It has been obtained in the best model, after an iterative
method to fit the observed relative emission-line intensities and the
corrected EW(H$\beta$) and L(H$\alpha$). In Fig. \ref{mod_lin} we show the
ratio between the intensities of the most representative
observed and modelled  emission lines for the three knots. Data for knot A
have been taken from \cite{Perez-Montero+10}. As we can see, the agreement
results excellent for all involved [\OIII] lines, with a deviation smaller than
5\% in all three knots. In the case of the [\OII] lines and [\SIII]\,9069\,\AA\ it
is better than 10\%. The fitting of [\SIII] at 6312\,\AA\ is a bit worse,
with a 20\% of disagreement in knots A and C, and 30\% in B.
The largest discrepancy is found for the [\SII] lines, from 30\% of
disagreement up to  65\% in the case of 6717,6731\,\AA\ in knot
B. In Table \ref{mod}, we compare the observed and modelled EW(H$\beta$),
corrected for the contribution of the underlying stellar population. We also give 
the number of ionizing photons and other properties predicted by
the individual models, such as the age of the ionizing cluster, filling
factor, ionization parameter, dust-to-gas ratio and visual extinction. Regarding
knot A, all quantities have been extracted from the  model in
\cite{Perez-Montero+10}. As we can see, the agreement between observed and
modelled values is excellent, both for the number of ionizing photons and the
EW(H$\beta$).

\begin{table}
\caption{Observed and model-predicted properties of the three studied regions}
\label{mod}
\begin{tabular}{@{}llccc@{}}
\hline
                  &           &      Knot A            &      Knot B            &      Knot C \\
\hline
Age (Myr)    &   &    7.9   &     5.1   &   4.6  \\
Abs. factor $f_d$  &   &   1.621  &   2.082   &   2.422  \\
log Q(H) (s$^{-1}$)  &  Obser. &   52.59  &   52.20  &  52.34  \\
			                        &  Model       &   52.59  &  52.26   &   52.29 \\
EW(H$\beta$)  (\AA)  & Obser.$^a$  &  132  &   153  &  167  \\
           		                    &	Model   &   127  &  158  &  164  \\
log Filling factor          &    &      -2.52  &  -2.02  &  -2.22  \\
log U                             &     &    -2.84  &   -2.54   &   -2.62 \\
log Dust-to-gas ratio          &     &   -1.96   &   -1.97  &  -1.81 \\
A$_V$                   &        &    0.32   &   0.49   &   0.55  \\

\hline
\multicolumn{5}{l}{$^a$Corrected for the underlying stellar population}
\end{tabular}
\end{table}

\begin{figure}
\centering
\includegraphics[width=8 cm]{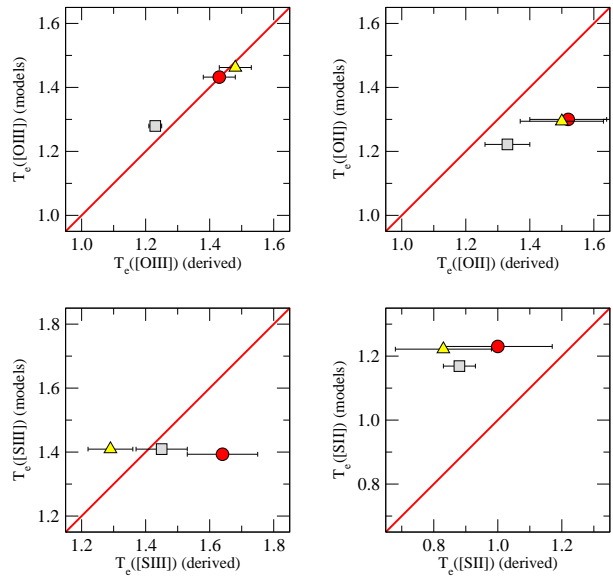}
\caption[]{Measured electron temperatures for the three knots, vs.\ values
  predicted by the photoionization models described in the text. The symbols
  are: grey squares, knot A; red circles, knot B, and yellow triangles,  knot
  C. Temperatures are in units of 10$^4$\,K.} 
\label{mod_temps}
\end{figure}

\begin{figure}
\centering
\includegraphics[width=8 cm]{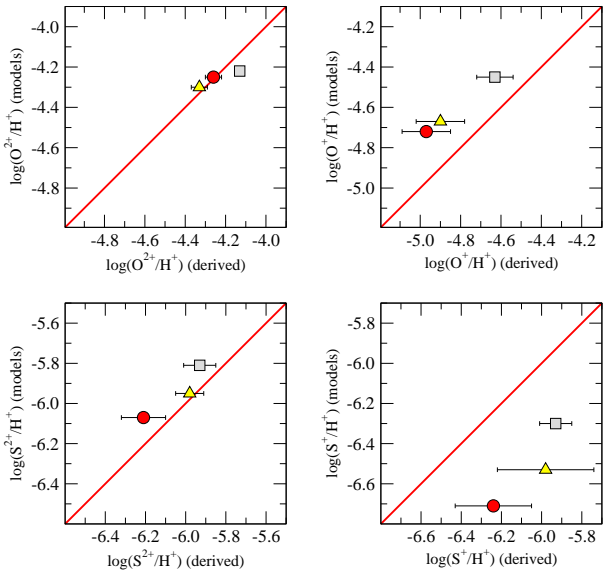}
\caption[]{Modelled vs.\ derived oxygen and sulphur ionic abundances (see
  description in the text). Knot A: grey squares; knot B: red circles; knot C:
  yellow triangles.}
\label{mod_ions}
\end{figure}

In Fig.\ \ref{mod_temps} we show a comparison between the four
measured electron temperatures in each of the three knots
and the values predicted by the models. As we can see, the best
agreement is found for T$_e$([\OIII]). In T$_e$([\SIII]),  a 
good agreement is found only for knot A. The model temperatures are higher for
T$_e$([\SII]) and lower for T$_e$([\OII]) than the derived from the measured
line intensities. In Fig.\ \ref{mod_ions}, we 
see the same comparison for the four respective ionic abundances.
{As in the case of electron temperatures, the agreement between O$^{2+}$
abundances derived from the observations and found by the models is excellent,
while in the case of O$^+$ and S$^{2+}$, only deviations not larger than 0.1
dex are found. The most evident deviation is found for the values of
S$^+$/H$^+$ which are higher in the direct measurements than in the model by
0.3 dex in average for the three knots.} 

We have calculated the total abundances of all the measured ions, now using
these models, taking into account when required, the unseen ionization stages
of each element, using the appropriate model predicted ICF for each species. 
The predicted ICFs for O, S, N, Ar and Ne and the  total abundances obtained
are listed in Table \ref{jhabicf}.  A discussion about the differences between
the ICFs calculated by these models and those obtained from the most commonly
used formulae is found in Appendix A of \cite{Perez-Montero+10}.

\section{Discussion}
\label{secDiscus}

\subsection{Gaseous physical conditions and element abundances}

\subsubsection{Densities and temperatures}

Four electron temperatures -- T$_e$([O{\sc iii}]), T$_e$([O{\sc ii}]),
T$_e$([S{\sc iii}]) and  T$_e$([S{\sc ii}])-- have been estimated for the
star-forming knots of \seisc. The good quality of the data allows us to reach
high precision, with rms errors of the order of 2\%, 5\%, 6\% and 6\%  in knot
A for T$_e$([O{\sc iii}]), T$_e$([O{\sc ii}]), T$_e$([S{\sc iii}]), and
T$_e$([S{\sc ii}]), respectively. For the faintest knots,  B and C, the
fractional errors are slightly higher, 3\%, 8\%, 6\% and 17\%, respectively.

The star-forming regions show temperatures within a relatively narrow range,
between 12000 and 14800\,K for T$_e$([O{\sc iii}]). It is worth remembering
that the adopted selection criteria for \seisc\  was
high H$\beta$ flux and large equivalent width of H$\alpha$, which tend to
render objects with abundances and electron temperatures close to the median
values shown by \HII galaxies. Although these criteria  applied 
to the main knot (the SDSS spectrum), we find similar electron temperatures
for all the regions. To our knowledge, there is no previously reported
T$_e$([O{\sc iii}]) for this galaxy in the literature. The estimated [O{\sc
    iii}] temperature for knot B is very similar to that of knot C; both are
higher than the T$_e$([O{\sc iii}]) for knot A by about 2000\,K.
{At same the time, although differences in T$_e$([S{\sc iii}]) among the
three knots are much larger, being this temperature 1900\,K larger in knot B
than in knot A, and 1600\,K lower in knot C than in knot A, these deviations
are still compatible within the errors with the empirical relation found in
Paper I between T$_e$([O{\sc iii}]) and T$_e$([S{\sc iii}]) for a
heterogeneous sample of Giant HII Regions and \HII galaxies.}

\subsubsection{Chemical abundances}

The abundances derived for the three knots  using the direct method
show the characteristic low values found in strong line \HII galaxies
\citep{Terlevich+91,Hoyos+06}. These values are in the range of
12+log(O/H)\,=\,7.78\,-\,7.99, in very good agreement with what is found from the
photoionization models discussed above, ranging  between 
\,7.80\,-\,8.02. The data presented in this paper is of 
high quality and the mean error values for the oxygen and
neon abundances are 0.05 dex, 0.12 for sulphur and 0.06 for argon. Knots B and
C show a similar value of 12+log(O/H), while Knot A is almost 0.2
dex higher. This difference is greater than the estimated
observational errors, and is similar (or even smaller) to what is found in
other works with spatial resolution of knots that belong to \HII galaxies or
Blue Compact Dwarf (BCD) galaxies \cite[see
  e.g.][]{Izotov+97,Vilchez+98,Kehrig+08,Cairos+09b,Perez-Montero+09,
Garcia-Benito+10}. However, in general, these differences were attributed to
the observational uncertainties (pointing errors, seeing variations, etc.) or
errors associated to the reddening correction and flux calibration, and the
oxygen abundance variations were not assumed as statistically significant,
concluding that there is a possible common chemical evolution scenario in all
of them. There are even greater differences when comparing the
estimated abundances of the individual knots with those derived from the
integrated spectra of the galaxies. For instance, \cite{Cairos+09b} found for
the integrated spectrum of Mrk\,1418 a lower value of direct oxygen abundance by
about 0.35\,dex (equivalent to a factor of 2.2) than for knots 1 and 2 of that
galaxy. They pointed out that while this variation could reflect a
real abundance difference in different scales (kpc-sized aperture for the
integrated spectrum and sizes of the order of 100\,pc for individual \HII
regions), it may also be due to  
relatively large measurement uncertainties for the weak [\OIII] auroral
emission line. Fortunately, our data are not affected by pointing errors and
seeing variations, and the other observational uncertainties have a second
order effect, since TWIN is a double beam longslit spectrograph that
simultaneously acquire all the observed spectral range. Likewise, the errors 
associated with the measurements of the weak auroral emission lines are
relatively small, specially for [\OIII].

The logarithmic N/O ratios found for \seisc\ using the direct method are
-1.23\,$\pm$\,0.11, -1.35\,$\pm$\,0.16 and -1.36\,$\pm$\,0.15 for knots A, B,
and C, respectively. The derived values are
on the high log(N/O) side of the distribution for this kind of objects (see
left-hand panel of Fig.\ 6 of Paper II).
The logarithmic values of this ratio found for the three knots using 
photoionization models  are slightly lower, -1.34\,$\pm$\,0.07,
-1.47\,$\pm$\,0.11 and -1.43\,$\pm$\,0.11, respectively, although similar
within the errors. However, the derived values present a larger uncertainty
than the values of total oxygen abundance, so a definite conclusion can not
be extracted about the homogeneity of this ratio. Anyway, it is quite
suggestive to find a larger N/O ratio in the brightest knot, which has the
larger metallicity. The N/O ratio is directly related to the chemical history
of galaxies, as these two elements have different nucleosynthetic origin, so
this difference can support to some extent the idea of a different chemical
evolution in the three knots of this galaxy.

\begin{figure*}
\centering
\includegraphics[width=.4\textwidth,angle=0]{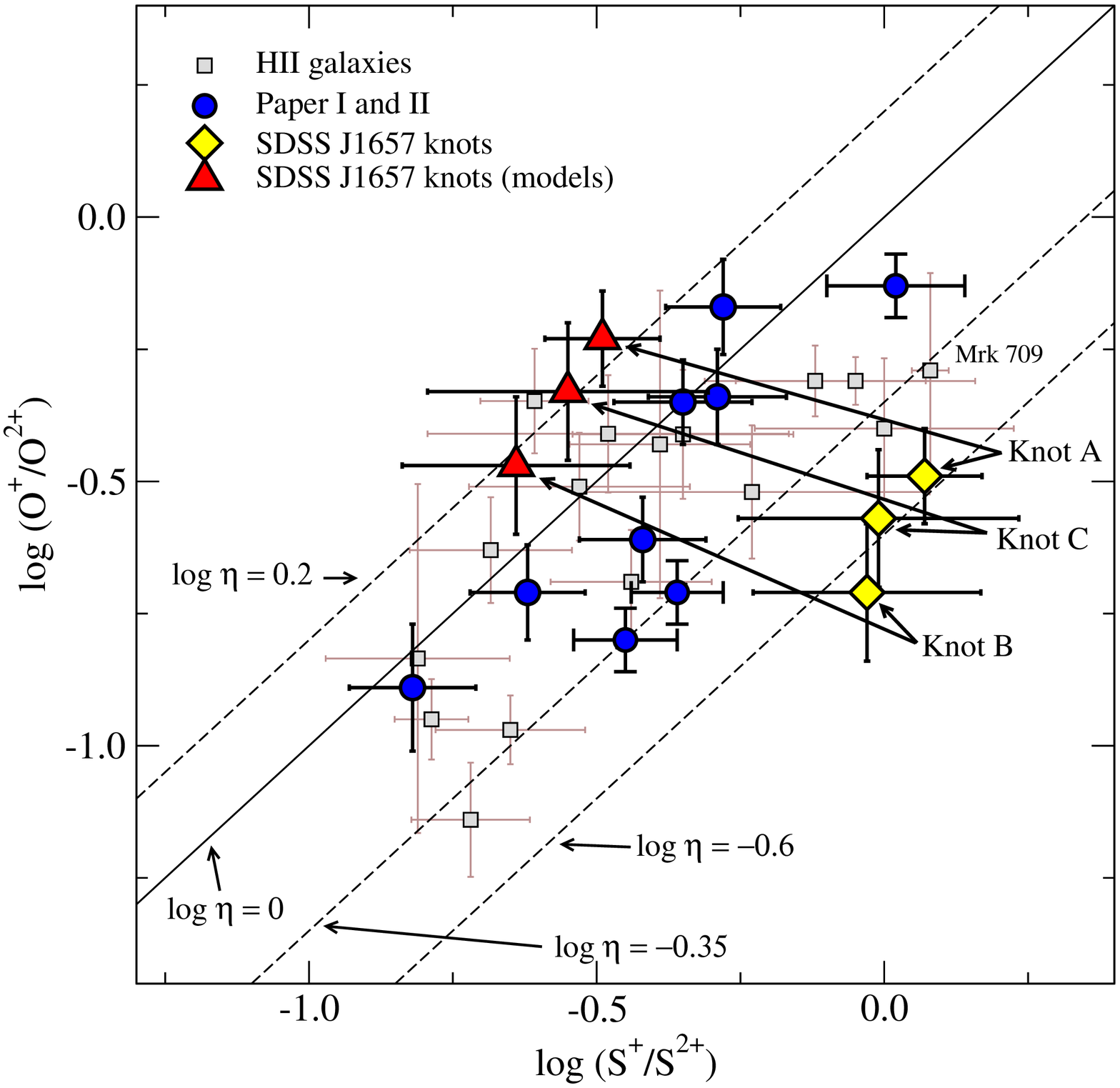}
\hspace{0.5cm}
\includegraphics[width=.4\textwidth,angle=0]{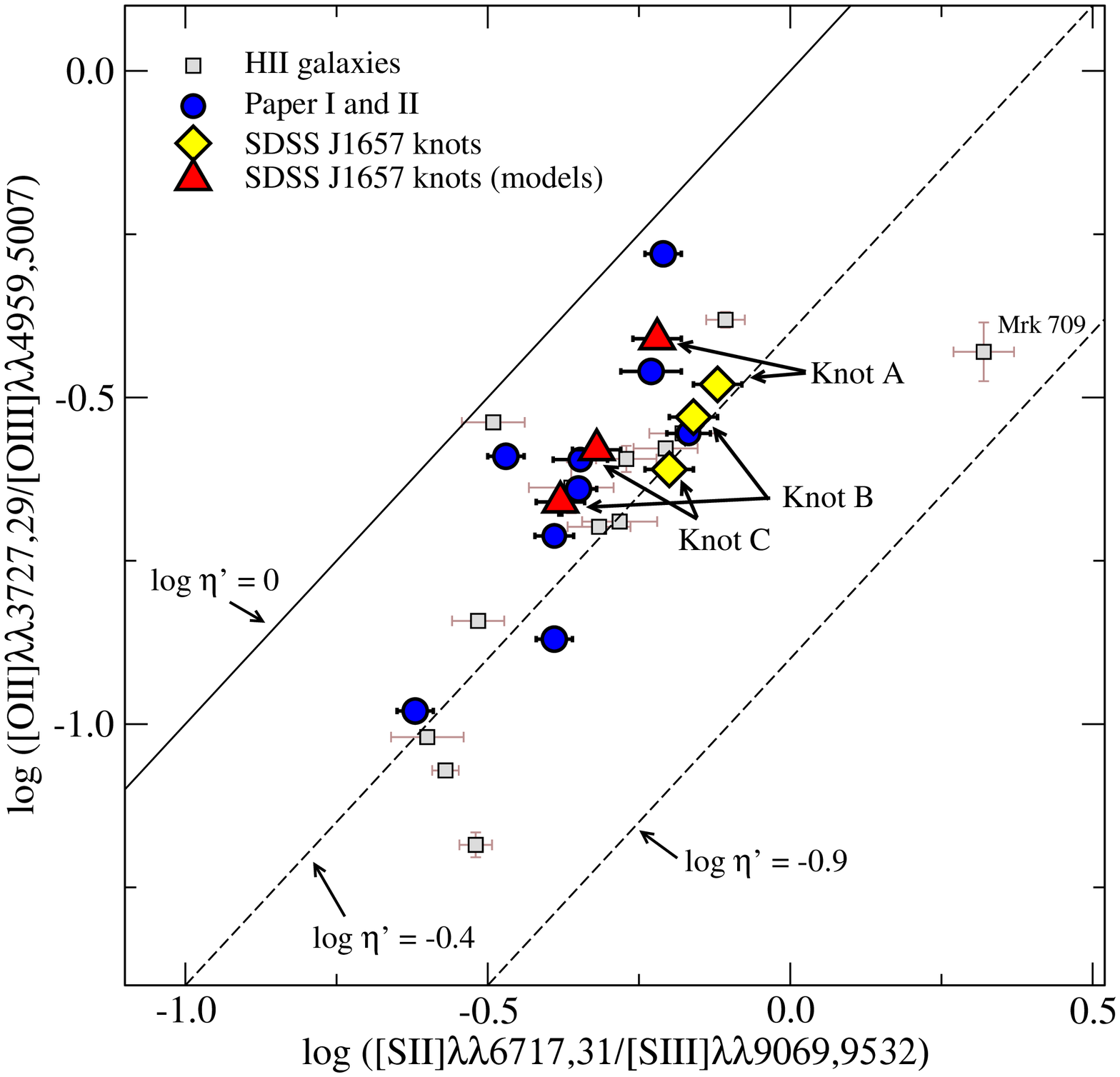}
\caption[]{Left panel: log(O$^+$/O$^{2+}$) 
vs.\ log(S$^{+}$/S$^{2+}$) using the direct method and the photoionization
models  (filled yellow diamonds and red triangles), the objects studied in
Paper I and Paper II (blue circles) and \HII galaxies from the literature
(open squares). Diagonals in this diagram correspond to constant
values of $\eta$. Right panel: log([O{\sc ii}]/[O{\sc iii}]) vs. log([S{\sc
ii}]/[S{\sc iii}]), symbols as in left panel. Diagonals in this diagram
correspond to constant values of $\eta$'.}
\label{eta}
\end{figure*}

The log(S/O) ratios found are  quite similar for Knots A and B
(-1.53 and -1.57 respectively) and higher for Knot C (-1.32), but all three
are almost equal within observational errors ($\sim$ 0.12 in average).
The values derived using the photoionization models follow
the same trend as those estimated using the
direct method (-1.65, -1.69, and -1.42
respectively). 
They are all slightly lower than the solar value, log(S/O)$_{\odot}$\,=\,-1.36 
\citep{Grevesse+98}.
On the other hand, the logarithmic Ne/O ratio is remarkably constant, with a
mean value of 0.75 (0.72 from the photoionization models), despite the differences in oxygen abundance
between knot A and knots B and C. They are consistent with solar,
log(Ne/O)\,=\,-0.61\footnote{Oxygen from \cite{Allende-Prieto+01} 
  and neon from \cite{Grevesse+98}.}.
The Ar/O ratios found (which are almost the same using the direct method and
the photoionization models) show a very similar value for Knot A and B,
while Knot C has a ratio higher by 0.2 dex. The mean value is consistent with
solar, log(Ar/O)\,=\,-2.29\footnote{Oxygen from \cite{Allende-Prieto+01} and
argon from \cite{Grevesse+98}.}, within the observational errors. 

Finally, the derived helium abundances are the same for the three knots within
observational errors. 

\subsection{Ionization structure}
\label{ionization}

The ionization structure of a nebula depends essentially on the shape of the
ionizing continuum and the nebular geometry and can be traced  by the ratio of
successive stages of ionization of the different elements. With our data it is
possible to use the O$^+$/O$ ^{2+} $ and the S$ ^{+} $/S$ ^{2+} $ to probe the
nebular ionization structure. In fact, \cite{1988MNRAS.231..257V} showed that 
the quotient of these two quantities that they called ``softness parameter" and
denoted by $ \eta $ is intrinsically related to the shape of the ionizing
continuum and depends on geometry only slightly. 
An insight into the ionization structure of the observed objects can be gained
by means of the $\eta$ diagram (see Paper I). 

In Fig.\ \ref{eta}, left panel, we show the relation between
log(O$^+$/O$^{2+}$) and log(S$^{+}$/S$^{2+}$) derived using the direct method
and the photoionization models for the knots of \seisc (filled
yellow diamonds and red triangles, respectively), the objects studied in Paper
I and Paper II (blue circles) and  \HII galaxies (open squares) from the
literature (see description and references in Paper II). In this 
diagram diagonal lines correspond to constant values of the $\eta$ parameter
which can be taken as an indicator of the ionizing temperature
\citep{1988MNRAS.231..257V}.  \HII\ galaxies occupy the region
with log\,$\eta$ 
between -0.35 and 0.2,  which corresponds to high values of the ionizing
temperature. As noticed in Paper II, Knot A
shows a very low logarithmic value of $\eta$, -0.6. Knots B and C
present very similar values. These objects, however, have the [O{\sc
ii}]\,$\lambda\lambda$\,7319,25 \AA\ lines affected by a set of atmospheric 
absorption lines. We treated  the problem (as is custom) by dividing the
spectra by a telluric standard. Unfortunately no previous data of this object
exists apart from the SDSS spectrum of Knot A. {While the agreement
between the [\OIII] lines measured in both the SDSS spectrum and ours is
good, the [\OII] lines measured on the SDSS spectrum provide a
T$_e$([\OII])\,=\,1.23\,$\pm$\,0.21, lower than the one derived from our
data.}  
This lower temperature would increase the value of O$^+$/O$^{2+}$ moving the
corresponding data point upwards in the left panel of Fig.\ \ref{eta}. 
This would be consistent with the position of the object in the
right panel of the figure showing log([O{\sc ii}]/[O{\sc iii}]) vs.\
log([S{\sc ii}]/[S{\sc iii}]), which does not require explicit knowledge of
the line temperatures involved in the derivation of the ionic ratios, and
therefore does not depend on the method to derive or estimate these
temperatures. The right panel in Fig.\ \ref{eta} shows the purely
observational counterpart of the left panel. In this diagram diagonal  
lines represent constant values of
log\,$\eta$'\,=\,log\,[([\OII]/[\OIII])/([\SII]/[\SIII])].  

\begin{figure*}
\centering
\includegraphics[width=0.95\textwidth,height=0.37\textheight]{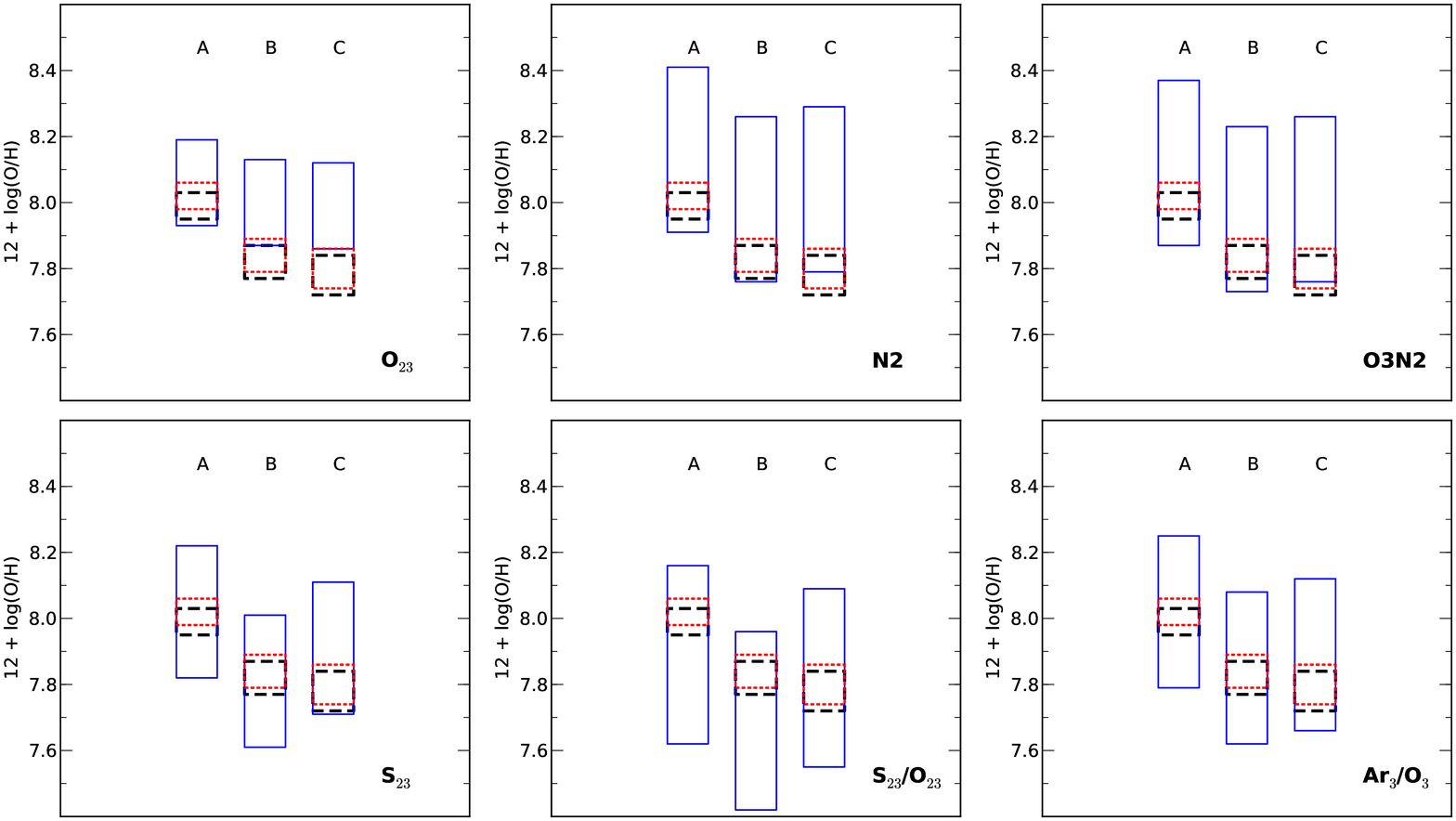}
\caption[]
{The oxygen abundances and their uncertainties for each observed knot of
\seisc (blue solid line rectangles), as derived using different
empirical calibrators. From left to right top panels: O$_{23}$, N2, and O3N2
and bottom ones: S$_{23}$, S$_{23}$/O$_{23}$, and Ar$_3$O$_3$. The
dash (black) and dotted (red) line rectangles represent the abundances and their
uncertainties as derived from the direct method and the photoionization
models, respectively.} 
\label{jhmet}
\end{figure*}

Another possible explanation for the differences between these two diagrams
can be obtained by inspecting the position of the models described above in
relation to the observational points (red triangles in the figures). Models
with a thick shell geometry and a constant density predict higher
T$_e$([\SII]) and, hence, lower S$^+$/H$^+$, which causes the $\eta$ parameter to be
higher than the values estimated from the measurements. 
In fact, the difference between the ionic abundances derived by the direct
method and those predicted by the models can be also seen in Fig.\
\ref{mod_ions}. 
On average there is a difference of 0.23\,dex for O$^+$/H$^+$, 0.07\,dex for
O$^{2+}$/H$^+$, 0.67\,dex for S$^{+}$/H$^+$ and 0.13\,dex for S$^{2+}$/H$^+$.
This effect, already pointed out by \cite{Perez-Montero+10},
could be a consequence of an outer shell of cold diffuse ionization structure
in these objects with an extra emission of [\SII] which contributes to their
integrated spectrum. 
The agreement is better for the $\eta$' diagram, being knot B the
most 
discrepant, which is consistent with the differences between the observed and modelled
lines (see Fig.\ \ref{mod_lin}), where the higher difference corresponds to
the [\SII] lines. 

In both diagrams, $\eta$ and $\eta$', the three knots of \seisc present a very
similar ionization structure, showing almost the same values within the
observational errors. This implies that the equivalent effective temperatures
of the ionization radiation field are very similar for all the
knots, although we find some small differences in the ionization state of the
different elements.

\subsection{Chemical abundances from empirical calibrators}

The emission line spectra of the three star-forming knots in J1657 are very
similar, implying similar values for ionization parameter, ionization
temperature, and chemical abundances. We derived the ionization parameters
from the [O{\sc ii}] to [O{\sc iii}] lines ratio according to the
expression given in \cite{Diaz+00b}. The logarithmic ratio is similar in all
the knots ranging from -2.47 for knot B to -2.65 for knot A.

The different strong-line empirical methods for abundance derivations, which
have been 
widely studied in the literature, are based on directly calibrating the
relative intensity of some bright emission lines against the abundance of some
relevant ions present in the nebula \citep[see e.g.][and references
  therein]{Garcia-Lorenzo+08,Cairos+09,RubenPhD,Garcia-Benito+10}. For the
case of oxygen, we take the 
calibrations studied by \cite{Perez-Montero+05}, who obtain different
uncertainties for each parameter in a sample of ionized gaseous nebulae with
accurate determinations of chemical abundances in the whole range of
metallicity. 

In Fig.\ \ref{jhmet}, we show the total abundances as derived
from several strong-line empirical methods (with their corresponding errors
estimated taking into account the errors of the line intensities and also the
errors given by the calibrations of the empirical parameters) and the oxygen
abundances calculated from the electron temperatures measured using the direct
method and those estimated from the photoionization models for each knot. 

Among the available strong-line empirical 
parameters we studied the O$_{23}$ parameter (also known as R$_{23}$ and
originally defined by \cite{Pagel+79} and based on [O{\sc ii}] and [O{\sc
iii}] strong emission lines). This parameter is characterized by its
double-valued relation with metallicity, with a very large dispersion in the
turnover region. According to the values measured, we used the
\cite{McGaugh91} calibration for the lower branch. For knots B and C, this
calibrator fails to predict the value obtained with the direct method,
overestimating the oxygen abundance, although the derived values are very
similar if we take into account the observational errors and the large spread
in the empirical O$_{23}$ diagram \citep[see Fig.\ 3 of][]{Perez-Montero+05}.

The N2 parameter \citep[defined by][]{Storchi-Bergmann+94} is based on the
strong emission lines of [N{\sc ii}]. It remains single-valued up to high
metallicities in its relation to oxygen abundance, and it is almost
independent of reddening and flux calibrations. Nevertheless, it has the high
dispersion associated to the functional parameters of the nebula (ionization
parameter and ionizing radiation temperature) and to N/O variations. We used
the empirical calibration of this parameter from \cite{Denicolo+02} to  
derive the oxygen abundance in the three star-forming knots of this galaxy. We
can see in Fig.\ \ref{jhmet} that N2 behaves similarly to O$_{23}$ in
predicting the abundances. 

The parameter O3N2, defined by \cite{Alloin+79}, depends on strong emission
lines of [O{\sc iii}] and [N{\sc ii}]. We used the calibration due to
\cite{Pettini+04} and, as we can see in Fig.\ \ref{jhmet} it has a very
similar behaviour to that of N2.

The S$_{23}$ parameter was defined by \cite{Vilchez+96} and is based on the
strong emission lines of [S{\sc ii}] and [S{\sc iii}]. The calibration by
\cite{Perez-Montero+05} yields comparable oxygen abundances  for the three
observed knots that are in turn in very good agreement with the abundances
derived using 
the direct method for knots A and B, and slightly higher for knot C, but still
consistent within the errors. 

The ratio of the S$_{23}$ and O$_{23}$ parameters as
S$_{23}$/O$_{23}$ \citep{Diaz+00a} is a parameter that increases
monotonically with the oxygen abundance up to the oversolar regime and  is
very useful to study variations over wide ranges of metallicity ({\em e.g.}
disks). We applied the calibration from \cite{Perez-Montero+05} and
found a good concordance with the values determined by the direct
method. 

The Ar$_3$O$_3$ parameter, defined and calibrated by \cite{Stasinska06} as the
ratio of [Ar{\sc iii}]\,$\lambda$\,7136\,\AA\ to [O{\sc
    iii}]\,$\lambda$\,5007\,\AA\ lines, 
predicts slightly higher values than the S$_{23}$ parameter and
in better agreement with those derived directly. 

Different strong line empirical metallicity calibrators are commonly used to
estimate the oxygen abundances in objects for which  direct derivation of
electron temperatures is not possible. 
However, as illustrated in Fig.\ \ref{jhmet}, different
empirical calibrations give results that are not in complete agreement with
direct measurements, and the goodness of the result even changes between knots
when using the same calibrator.  \cite{Perez-Montero+09} found a very similar
behaviour for 
the star-forming knots of IIZw71, except for the Ar$_3$O$_3$ parameter, which
is a better estimator of the oxygen abundances for the knots in
\seisc. For two knots in Mrk\,1418, \cite{Cairos+09b} derived the oxygen 
abundances from the observed [\OIII]\,$\lambda$\,4363\,\AA\ auroral emission
line, and compared them with those abundance values derived using the N2 and
O3N2 empirical parameters, and found a similar discrepancy, with the
empirically derived values being slightly larger.
Clearly, more observations and direct abundance estimations are needed in
order to improve calibrations and truly understand the origin of the observed
dispersions and discrepancies (see discussion in Paper I about the dispersion
and precision of the S$_{23}$ parameter).

\subsection{The stellar population}
\label{stpop}

The study of the stellar content of our objects was
carried out using the STAR\-LIGHT code, which calculates the combination of
stellar libraries and the extinction law that reproduces the spectral energy
distribution, to derive the properties of the stellar  
population in each of the knots. STARLIGHT fits an observed continuum spectral
energy distribution using a combination of multiple simple stellar populations
(SSPs; also known as instantaneous bursts) synthetic spectra using a $\chi^2$
minimization procedure. 
For consistency with the photoionization models used to model
the gas of knot A in  \cite{Perez-Montero+10}, we have used in the
fitting a synthetic stellar population obtained using Starburst99
\citep{1999ApJS..123....3L,2005ApJ...621..695V} with the Geneva stellar
evolutionary tracks for continuous star formation with 
high mass loss \citep{1994A&AS..103...97M}, the Kroupa Initial Mass Function
\citep[IMF;][]{2002Sci...295...82K} in two intervals (0.1-0.5 and
0.5-100\,M$_\odot$) with different exponents (1.3 and 2.3, respectively), the
theoretical wind model \citep{1992ApJ...401..596L}, with the model
atmospheres from \cite{2002MNRAS.337.1309S}, and the stellar cluster
metallicity being the closest to the nebular one, Z = 0.004
\cite[=\,1/5\,Z$_\odot$, see Paper II and][]{Perez-Montero+10}.
The STARLIGHT code solves simultaneously the ages and relative contributions
of the different SSPs and the average reddening. The reddening law from
\cite{Cardelli+89} is used. Prior to the fitting procedure, the spectra
were shifted to the rest-frame, and re-sampled to a wavelength interval of 1
{\AA} in the entire wavelength range between 3500 {\AA} and 9000 {\AA} by
interpolation conserving flux, as required by the program. Bad pixels and
emission lines were excluded from the final fits. 

\begin{table}
\caption[]
{Values of the extinction, total stellar mass, and fraction of the mass in
stars younger than 10 Myr for each knot of \seisc\ in the best-fit model
using STARLIGHT.} 
\label{mass}
\begin{tabular}{@{}l c c c c c@{}}
\hline
\multicolumn{1}{c}{ID}    & A(V)    & c(H$\beta$)  &  M$_*$  & M$_{ion}$       & f(age $<$ 10 Myr) \\
                          & (mag)   &              &  &  & (\%) \\
\hline
Knot A      &  0.18  &  0.08         & 280  & 1.6  & 0.58   \\ 
Knot B      &  0.00  &  0.00         & 100  & 0.2  & 0.21   \\  
Knot C      &  0.00  &  0.00         & 140  & 0.4  & 0.28  \\ 
\hline
\multicolumn{5}{@{}p{0.2\textwidth}}{Note. Masses in 10$^6$\,M$_\odot$.}
\end{tabular}
\end{table}

\begin{figure}
\centering
\includegraphics[width=0.45\textwidth,clip=]{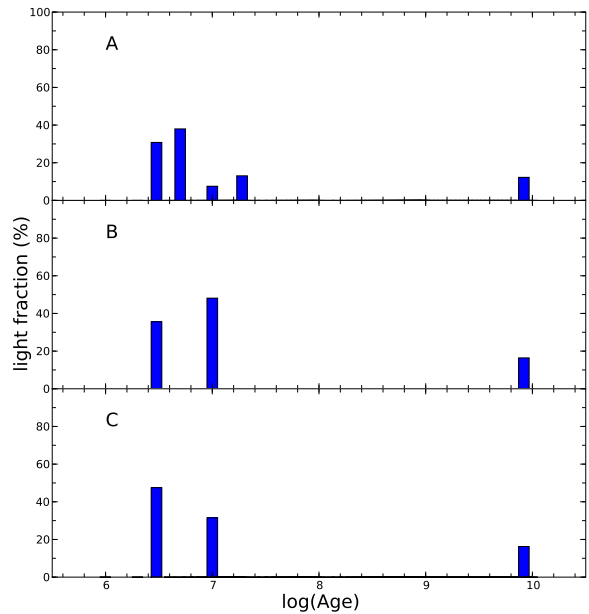}
\caption[]
{Histogram of the distribution in visual light of the most probable stellar
  population models fitted by STARLIGHT for  knots A, B and C, as labelled.}
\label{jhsl}
\end{figure}

In Fig.\ \ref{jhsl} we show the age distribution of the visual light
fraction for the individual knots. All of them present a very young stellar
population with ages around 10 Myr, responsible for the ionization of the
surrounding gas. Practically all the mass of the knots seems to come
from a very old population of about 8.3 Gyr.  However, this result is puzzling
given that no absorption metal lines characteristic of old stellar
populations, such as Mg{\sc ii}\,$\lambda$\,5173\,\AA\ and/or Ca{\sc
  ii}\,H\,\,$\lambda$\,3933\,\AA, Ca{\sc ii}\,K\,$\lambda$\,3968\,\AA, and
Ca{\sc ii} Triplet $\lambda$$\lambda$\,8494, 8542, 8662\,\AA,  
are detected in the spectra. The estimated total stellar mass, the mass of the
stellar population with an age younger than 10 million years, responsible for
the ionization of the gas, and the fraction of this last one with respect to
the former, are listed in Table 
\ref{mass} for the three knots, together with the internal extinction,
 A(V) (as given by STARLIGHT) and the reddening constant [c(H$\beta$)]
estimated by the model. No aperture correction factors have been taken into
account for the H$\alpha$ luminosities, due to the compact nature of the
objects. Indeed, the discrepancy factors between our H$\alpha$ luminosities for
Knot A and those measured in the SDSS catalog using a 3 arcsec fiber is no
larger than 1.3. 

We do not expect to find the same values of extinction in the gas and the
stellar population. In  this galaxy, although the extinction is
larger in knot A than in B and C (the opposite from the
derived values using the Balmer decrement) the general result is
consistent with low extinction. 

The mean ages and stellar mass fraction (with respect to the total
stellar mass of the corresponding cluster) of the ionizing stellar population
fitting by STARLIGHT for each knot is very similar, in very good agreement
with the ionization structure derived in \S\ref{ionization}. Since the
age distribution of the ionizing population of the different knots 
seems to be similar, this could indicate a common evolutionary stage of
this population which is probably related to a process of interaction with a
companion galaxy that triggered the star formation in the different knots
almost at the same time. In Fig.\ \ref{imgalknots} we can appreciate a
tail of diffuse emission gas observed to the south of knot A, with a few weak
and blue knots  probably hosting star formation. To the north-east there is a
bright and redder galaxy. We need more spectroscopic data to disentangle
whether these different stellar systems are related or it is only a projection
effect. The relatively small differences in the derived metallicities for the
star-forming knots of \seisc could indicate a slightly different  chemical
history of these knots. 

\subsection{Ionizing stellar populations}

Some properties of the emission knots can be obtained from the measured
H$\alpha$ flux, such as H$\alpha$ luminosity, number of ionizing photons, mass
of ionizing stars and mass of ionized hydrogen \cite[see
  e.g.][]{Diaz+00b}. The observed H$\alpha$ flux was corrected in each knot
for reddening using the values of the reddening constants, c(H$\beta$), given
in Table \ref{ratiostot}. We assume a distance to \seisc of 161.2\,Mpc
\citep{2000ApJ...529..786Mtot}. 
Besides, we have used the dust absorption factors, $f_d$ derived using 
photoionization models to correct these quantities. These factors are
independent of reddening correction because they affect above all the
number of ionizing photons emitted by the stellar cluster.  As we can see in
Table \ref{mod}, these factors are larger in knots B and C than in knot A,
consistent with the differences found in the reddening as derived by measuring
the Balmer decrement. These differences in reddening are not surprising
attending to the different ages of the ionizing clusters found in the same
photoionization models (see \S\ref{photomod}), which are respectively 7.9, 5.1
and 4.6. Although the metallicity in 
knots B and C is lower, as the star formation started later in these knots, 
the ionizing photons have not already broken the dust cocoon usually shrouding
these very young star formation knots. The derived and corrected values are
given in Table \ref{prop}.  

\begin{table}
\centering
\caption[Derived properties of the observed knots from the H$\alpha$ fluxes]
{Derived properties of the observed knots using the extinction and dust
absorption corrected measured H$\alpha$ fluxes.}
\label{prop}
\begin{tabular}{@{}lcccccc@{}}
\hline
\multicolumn{1}{c}{ID} & F(H$\alpha$) & L(H$\alpha$) & Q(H$_0$) & M$_{ion}$ & M(\HII)  & SFR \\
\hline
Knot A  & 28.4 & 8.3 & 6.3 & 3.6 & 2.1 & 0.697 \\
Knot B  & 15.2 & 4.8 & 3.3 & 1.5 & 3.5 & 0.375 \\
Knot C  & 23.0 & 7.0 & 5.3 & 2.2 & 5.3 & 0.564 \\
\hline
\multicolumn{7}{@{}p{0.46\textwidth}}{Note. Fluxes in
10$^{-15}$\,erg\,s$^{-1}$\,cm$^{-2}$, luminosities in
10$^{40}$\,erg\,s$^{-1}$, ionizing photons in 10$^{52}$\,photons\,s$^{-1}$,
masses in 10$^6$\,M$_\odot$, and SFR in M$_{\odot}$\,yr$^{-1}$.}
\end{tabular}
\end{table} 

The derived values of the masses of the ionizing clusters ({\em i.e.} with an
age younger than 10 Myr) can be compared with
those provided by the STARLIGHT fit. For knot A STARLIGHT gives
a slightly lower value than the one we derive by a factor of 2, while for knot
B and C this factor is larger, about 7 and 5, respectively. If we take the
masses derived from the H$\alpha$ fluxes, which are, in principle, lower
limits since a possible escape of photons is not taken into
account, and we use the calculated proportions of young to total mass given in
Table \ref{mass} we obtain total masses of about 7\,$\times$\,10$^8$
M$_{\odot}$ for each knot.

The weight of the underlying stellar population in the visual light 
is larger in knot A than in knots B and C. The
corrected  EW(H$\beta$) is 12\% larger in knot A but
less than 1\% in the other two knots.

The star formation rate (SFR) for each knot was derived from the H$\alpha$
luminosity using the expression given by \cite{Kennicutt98},
SFR\,=\,7.9\,$\times$\,10$^{-42}$\,$\times$\,L(H$\alpha$). 
The derived values are also given in Table \ref{prop}, taking into account
dust absorption factors obtained from our models as discussed in previous
sections. The SFR obtained range  from 0.375 M$_\odot$ yr$^{-1}$ for knot B to
0.697\,M$_\odot$\,yr$^{-1}$ for knot A.

\subsection{Kinematics of the knots}

\begin{figure}
\centering
\includegraphics[bb=0 3 57 36,width=0.45\textwidth,clip=]{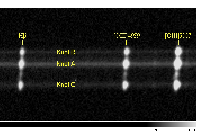}\\
\vspace{0.1cm}
\includegraphics[bb=0 3 57 36,width=0.45\textwidth,clip=]{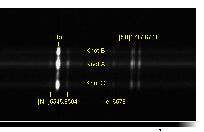}\\
\vspace{0.2cm}
\includegraphics[width=0.43\textwidth,clip=]{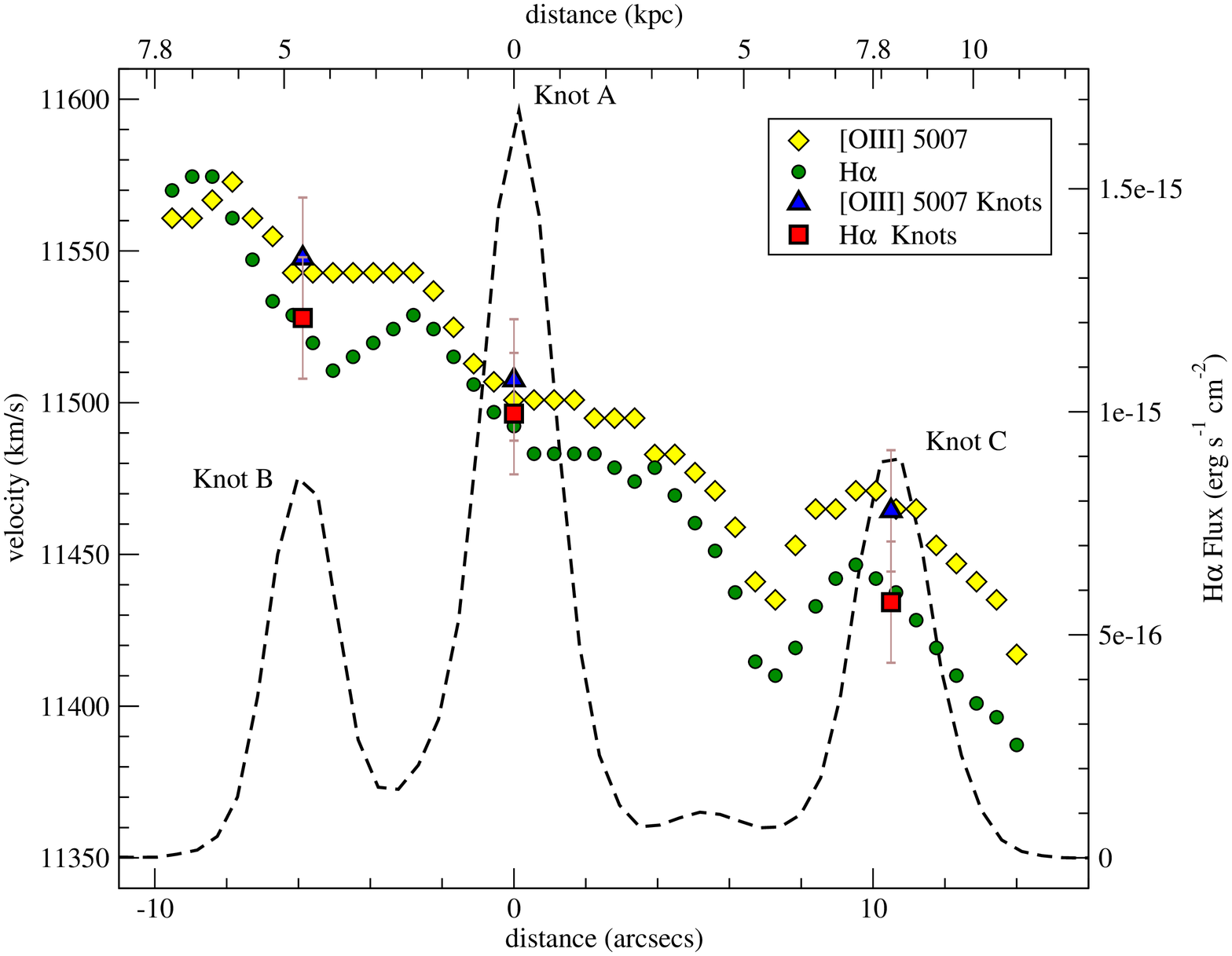}
\caption[]
{Longslit spectrum between 5030 and 5220\,\AA\ (upper panel) and between 6690
  and 7110\,\AA\ (middle panel);  the main emission lines
and the knots are labelled. The derived rotation
  curve of \seisc  is shown in the lower panel. Green circles
  represent the radial velocities derived from H$\alpha$ and yellow diamonds
  from [\OIII]\,5007\,\AA\ for each pixel (0.56 arcsec) along the slit. Red
  squares and blue upward triangles represent the radial velocities derived from
  H$\alpha$ and [\OIII] using the extracted spectrum for each knot. The dashed
  line represents the spatial distribution of the H$\alpha$ flux to be compared
  with the rotation curve.} 
\label{kine}
\end{figure}

In Fig.\ \ref{kine} we show the longslit spectrum between 
5030 and 5220\,\AA\ (upper panel) and between 6690 and 7110\,\AA\ (middle
panel). These spectral ranges contain  H$\beta$ and
[\OIII]\,4959,5007\,\AA, and H$\alpha$, [\NII]\,6548,6584\,\AA,
\HeI\,6678\,\AA\ and [\SII]\,6717,6731\,\AA\ emission lines, respectively. All
these lines show a similar behaviour in the
position-velocity (wavelength) plane, with the typical helical shape of the 
rotation curve expected for a rotating disc. Knot B is redshifted while Knot C
is blueshifted with respect to the systemic velocity of knot A. We
have analyzed the differential radial velocity field along the slit using the
\Ha and [\OIII] emission lines. The radial velocities derived for each pixel
(0.56 arcsecs) are displayed in the lower panel of Fig.\ \ref{kine}. We have
also plotted the \Ha spatial profile. The  velocities derived from both
emission lines are in good agreement, giving the same value within the
observational errors, which are about 
20\,km/s, taking into account the errors in the wavelength calibration and a
single Gaussian fitting. Larger differences ($\sim$\,30\,km/s) between the 
derived velocities using the different emission lines have been found around
the secondary knots, B and C, but they are also within the observational
errors. From the data for Knot A, we have estimated a redshift of 0.0383
(v$_r$\,$\approx$\,11500\,km/s), in very good agreement with the value given
by the SDSS (see Table \ref{obj}).

The rotation curve is basically linear, with some perturbations. Also the weak
pure emission line knot located between Knot A and C seems to follow a
circular rotation curve. There is a depression of about 50\,km/s in the radial
velocity curve located between this pure emission knot and Knot C. In
principle, we could attribute this deviation from the circular motion to a poor 
signal from the emission lines (see the spatial profile and the
longslit spectra, Figs.\ \ref{profiles} and \ref{kine},
respectively). However, it is not the position with 
the lowest emission, and the other zone around the minimum emission does not
show a similar deviation of the radial velocity field from the circular
motion. Moreover, the radial velocities in this depression derived from the \Ha and
[\OIII]  lines are in good agreement showing a similar behaviour,
which leads us to suppose that this depression is a real deviation from
the circular motion. This could be due to the presence of an expanding bubble
or shell (or superposition of more than one) of ionized gas approaching us
with respect to the systemic velocity of the galaxy defined by the velocity of
the brightest and most massive star-forming cluster, Knot A. In this case, we
could only see the expanding material which is moving in our direction, since
we detect a systematic variation of the radial velocity, but not a broadening
of the emission line with a systemic velocity in accordance with that expected
for the circular motion. Another possibility would be that the ionized gas
related to this weak emission area
is affected by a possible interaction with a tail of emission gas located at
the south of the Knots or with the redder galaxy observed to the north-east
(see Fig.\ \ref{imgalknots} and discussion above). This interaction could be
responsible for the deviation from circular motion,
as well as the fact that the ionizing population of the different star-forming
knots seem to have similar ages (see \S\ref{stpop}). It has been noticed by
\cite{Cairos+09b} that several works show
a substantial fraction of dwarf galaxies that show recent episodes of star
formation have optical faint and gas-rich low-mass companion galaxies
\citep[e.g.][]{Taylor+93,Taylor+95,Taylor+96,Noeske+01,Pustilnik+01}. Even in
the case of apparent isolated galaxies there are clues that favour the
interpretation that interactions play a substantial role in the evolution of
these individual systems
\citep{Johnson+04,Ostlin+04,Bravo-Alfaro+04,Bravo-Alfaro+06,Cumming+08,Cairos+09b}.

Assuming that we see the galaxy edge-on, with a radial velocity
v$_r$\,=\,100\,km/s given by the difference between Knot A and the furthest
point where we can measure the emission lines with good enough S/N,
and considering an optical radius of 14\,arcsec for this point (equivalent to
R$_{\textrm{opt}}$\,=\,11\,Kpc at the adopted distance of 161.2\,Mpc for
\seisc, with a scale of 782\,pc\,arcsec$^{-1}$) we have
estimated a dynamical mass inside this radius of
2.5\,$\times$\,10$^{10}$\,\Msun. 

Since the three star-forming knots clearly dominate the luminosity of the
galaxy (see Fig.\ \ref{imgalknots}), we have derived their blue magnitudes
from the SDSS values (see Table \ref{obj}). We adopt these values as
representative of the total 
light, although we know that they are (together) an upper limit to the total
magnitude of the galaxy since we are not taking
into account the light outside the 3 arcsec SDSS fiber (see Fig.\
\ref{imgalknots}). We have obtained the B magnitude from the Sloan g and r
photometric magnitudes using the transformation equation from
\cite{Chonis+08}, B\,=\,g+0.327$\times$(g-r)+0.216, given B$_{\textrm
A}$=17.15, B$_{\textrm B}$=20.75, and B$_{\textrm C}$=19.67. We have obtained
a total blue luminosity L$_{\textrm B}$\,=\,6.2$\times$10$^9$\,L${\sun}$,
which gives an upper estimation of 4 for  M/L$_{\textrm B}$, in agreement
to what is found by \cite{Faber+79} for irregular galaxies.

\section{Summary and conclusions}
\label{secConclus}

The star formation processes in \HII galaxies are known to occur in low density
environments. 
Spectrophotometric observations of \seisc were carried out to study the
physical properties of its individual bursts of star formation. In the three
individualized star-forming regions the electron densities were 
found to be well below the critical density for
collisional de-excitation. 

We extracted information on the three knots of \seisc, labelled  A, B and 
C. For all of them  we measured four line temperatures:
T$_e$([\OIII]), T$_e$([\SIII]), T$_e$([\OII]), and T$_e$([\SII]), reaching
high precision, with rms fractional errors  of the order of 2\%, 5\%, 6\% and 6\%,
respectively, for Knot A, and  slightly worse 3\%, 8\%, 6\% and 17\%,
respectively, for the fainter knots B and C.
The [\OIII] temperature of  knot A was
found to be about 2000 K lower than for the other two knots. For the [\OII]
temperature the estimated values for the three knots are the same within the
errors. The [\SIII] temperatures show a difference of 3500\,K between the
values derived for Knots B and C, and the estimated value for Knot A falls in
between. {However, these large differences are compatible with the
empirical relation between T$_e$([\OIII]) and T$_e$([\SIII]) found in Paper
I, within the observational errors and the empirical dispersion of that
relation.} Within observational errors, the [\SII] temperature derived is
similar for the three knots. 

The temperature measurements allowed the direct derivation of ionic abundances
of oxygen, sulphur, nitrogen, neon and argon. The total abundances of these
species are in the same range of metallicities measured in \HII galaxies,
between 7.78 and 7.99, with estimated errors of about 0.05\,dex. Knots B and C
show similar abundances, while the value for Knot A  is about 0.2\,dex
higher. This behaviour is mirrored by the N/O ratio. Knots A and B have
similar Ar/O while Knot C is the one with a different 
value. Within obsevational errors, S/O  is almost the same for the three knots.
The Ne/O ratio is remarkably similar
for all three regions.

We have studied the underlying and ionizing stellar populations for the three
regions and  modelled the properties of the emitting gas using a
photoionization code. The estimated electron temperatures and ionic abundances
using the direct method are well reproduced by the models except for those of
S$^+$/H$^+$ . This could be due to the presence of diffuse gas in these
star-forming regions, which is not taken into 
account in the models. Regarding the hardness of the radiation field, model
predictions agree with observations when the softness parameter $\eta$ which
parametrizes the temperature of the ionizing radiation, is expressed in terms
of emission lines intensity ratios. However, there is a disagreement when ionic
abundance ratios are used instead. This is probably caused by the overestimate
of the electron temperature of S$^+$ by the models and the corresponding
underestimate of the S$^+$/S$^{2+}$ abundance ratio rather than by the existence
of an additional heating source in these knots.

We have also estimated the total oxygen abundances by means of different
strong-line empirical parameters. In all cases, the estimated abundances are
consistent with those derived by the direct method with the parameter
involving the sulphur lines providing the closest fit. The rest of
the parameters slightly overestimate the oxygen abundance.

The star formation history for the three knots derived from the fitting of
multiple simple stellar populations using STARLIGHT are remarkably similar,
with an old population of about 8 Gyr presenting more than the 99\%
contribution to the mass fraction. This result is somewhat unexpected 
and data on its surface brightness distribution would be very valuable to
explore further this finding. Since the
age distributions of the ionizing population among the different knots of
\seisc seem to be similar, this could indicate a common evolutionary stage of
this population which is probably related to a process of interaction with a
companion galaxy that triggered the star formation in the different knots
at about the same time. Besides, this interaction could be responsable for
the deviation from the circular motion shown by the rotation curve, which is
basically linear in the region where we can measure the emission lines with a
good enough S/N ratio. However, this deviation could be also related to an
expanding bubble or shell of ionized gas approaching us with a velocity of
50\,km/s with respect to the predicted velocity from the rotation curve.

The ionization structure mapped through the use of the $\eta$ and $\eta'$
diagrams derived from our observations shows very similar values within the
errors for all the knots. This fact implies that the
equivalent effective temperatures of the ionization radiation
fields are very similar for all the studied regions, in spite of
some small differences in the ionization state of different
elements.

Finally, we have derived a dynamical mass of 2.5\,$\times$\,10$^{10}$\,\Msun
from the rotation curve with a mean value of 100\,km/s at an radius of
11\,Kpc.
The total mass of the young clusters derived for the three star-forming
knots using the H$\alpha$ luminosities is 7.3\,$\times$\,10$^{6}$\,\Msun,
making up a small fraction of the total dynamical mass, about 0.03\,\%. 
We have estimated an upper limit of about 4 for the ratio M/L$_{\textrm
  B}$. Data on the surface brightness 
distribution using broad and narrow band images would be very valuable to
explore further this result.

\section*{Acknowledgments}

We acknowledge fruitful discussions with Yago Ascasibar and Amalia Meza. We
also thank very much an anonymous referee for his/her careful examination
of our manuscript. 

We would like to thank the time allocation committee for awarding time to this
project and the support staff at Calar Alto for their cheerful and valuable
assistance. 

Funding for the creation and distribution of the SDSS Archive has been
provided by the Alfred P. Sloan Foundation, the Participating Institutions,
the National Aeronautics and Space Administration, the National Science
Foundation, the US Department of Energy, the Japanese Monbukagakusho, and the
Max Planck Society. The SDSS Web site is http://www.sdss.org.

The SDSS is managed by the ARC for the Participating Institutions. The
Participating Institutions are the University of Chicago, Fermilab, the
Institute for Advanced Study, the Japan Participation Group, The Johns Hopkins
University, the Korean Scientist Group, Los Alamos National Laboratory, the
Max Planck Institute for Astronomy (MPIA), the Max Planck Institute for
Astrophysics (MPA), New Mexico State University, the University of Pittsburgh,
the University of Portsmouth, Princeton University, the United States Naval
Observatory, and the University of Washington.

This research has made use of the NASA/IPAC Extragalactic Database (NED) which
is operated by the Jet Propulsion Laboratory, California Institute of
Technology, under contract with the National Aeronautics and Space
Administration and of the SIMBAD database, operated at CDS, Strasbourg,
France. 

Financial support for this work has been provided by the Spanish
\emph{Ministerio de Educaci\'on y Ciencia} (AYA2007-67965-C03-03 and
02). Partial 
support from the Comunidad de Madrid under grant S2009/ESP-1496 (ASTROMADRID)
is acknowledged. EPM also thank to project Junta de Andaluc\'ia TIC 114. ET
and RT are grateful to the Mexican Research Council 
(CONACYT) for support under grants CB2005-01-49847F  and CB2008-01-103365. 
VF would like to thank the hospitality of the Astrophysics Group of
the UAM during the completion of this work.

 \bibliographystyle{mn2e}
 \bibliography{references}

\end{document}